\documentclass[a4paper,11pt]{article}
\pdfoutput=1 
\usepackage{jheppub} 
\usepackage[T1]{fontenc} 
\usepackage{tikz}
\usepackage{footnote}
\usepackage{slashed}
\usepackage{tablefootnote}
\usepackage{multirow}
\usepackage{ulem}
\usetikzlibrary{arrows.meta}
\tikzset{%
  >={Latex[width=2mm,length=2mm]},
            base/.style = {rectangle, rounded corners, draw=black,
                           minimum width=4.5cm, minimum height=1.4cm,
                           text centered, font=\sffamily},
  activityStarts/.style = {base, fill=white}
}

\usepackage{makecell}

\def\simge{\mathrel{%
   \rlap{\raise 0.511ex \hbox{$>$}}{\lower 0.511ex \hbox{$\sim$}}}}
\def\simle{\mathrel{
   \rlap{\raise 0.511ex \hbox{$<$}}{\lower 0.511ex \hbox{$\sim$}}}}
\def\s#1{\setbox0=\hbox{$#1$}%
\rlap{\ifdim\wd0>.7em\kern.22\wd0\else\kern.1\wd0\fi /}#1}

\newcommand{\newc}{\newcommand}

\newc{\ol}{\overline}
\newc{\wt}{\widetilde}
\newc{\bs}{\boldsymbol}
\newc{\m}{\mathcal}
\newc{\la}{\lambda}
\newc{\lra}{\longrightarrow}
\newc{\vp}{\varphi}
\newc{\ti}{\tilde}

\preprint{TTP21-048, P3H-21-093, LAPTH-043/21}

\title{\boldmath Data-driven analysis of a SUSY GUT of flavour}

\author[a,b]{Jordan~Bernigaud}
\author[c]{~~Adam~K.~Forster}
\author[d]{~~Bj\"orn~Herrmann}
\author[c]{~~Stephen~F.~King}
\author[e]{~~Werner Porod}
\author[c]{~~Samuel~J.~Rowley}

\affiliation[a]{Institute for Nuclear Physics (IKP), Karlsruhe Institute of Technology, Hermann-von-Helmholtz-Platz 1, D-76344 Eggenstein-Leopoldshafen, Germany}
\affiliation[b]{Institute for Theoretical Particle Physics (TTP), Karlsruhe Institute of Technology, Engesserstrasse 7, D-76128 Karlsruhe, Germany}
\affiliation[c]{School of Physics and Astronomy, University of Southampton, SO17 1BJ Southampton, United Kingdom}
\affiliation[d]{Univ.\ Grenoble Alpes, Univ.\ Savoie Mont Blanc, CNRS, LAPTh, F-74000 Annecy, France}
\affiliation[e]{Institut f\"ur Theoretische Physik und Astrophysik, Campus Hubland Nord, Univ.\ W\"urzburg, Emil-Hilb Weg 22, D-97074 W\"urzburg, Germany}

\emailAdd{jordan.bernigaud@kit.edu}
\emailAdd{a.k.forster@soton.ac.uk}
\emailAdd{herrmann@lapth.cnrs.fr}
\emailAdd{s.f.king@soton.ac.uk}
\emailAdd{porod@physik.uni-wuerzburg.de}
\emailAdd{s.rowley@soton.ac.uk}

\abstract{We present a data-driven analysis of a concrete Supersymmetric (SUSY) Grand Unified Theory (GUT) of flavour, based on $SU(5)\times S_4$, which predicts charged fermion and neutrino mass and mixing, and where the mass matrices of both the Standard Model and the Supersymmetric particles are controlled by a common symmetry at the GUT scale. This framework also predicts non-vanishing non-minimal flavour violating effects, motivating a sophisticated data-driven parameter analysis to uncover the signatures and viability of the model. This computer-intensive Markov-Chain-Monte-Carlo (MCMC) based analysis includes a large range of flavour as well as dark matter and SUSY observables, predicts distributions for a range of physical quantities which may be used to test the model. The predictions include maximally mixed sfermions, $\mu\rightarrow e \gamma$ close to its experimental limit and successful bino-like dark matter with nearby winos (making direct detection unlikely), implying good prospects for discovering winos
and gluinos at forthcoming collider runs. 
The results also demonstrate that the Georgi-Jarlskog mechanism does not provide a good description of the splitting of down type quark masses and charged leptons, while neutrinoless double beta decay is predicted at observable rates.}

\begin{document} 
\maketitle
\flushbottom

\section{Introduction}
\label{sec:intro}

The Minimal Supersymmetric Standard Model (MSSM) remains an appealing extension of the Standard Model of particle physics, as it provides solutions to the most prominent shortcomings of the latter. In addition to solving the hierarchy problem related to the mass of the Higgs boson \cite{HiggsATLAS2012, HiggsCMS2012}, the model includes a viable candidate for the observed Cold Dark Matter (CDM) in the Universe, namely the lightest of the four neutralinos. Furthermore, the masses of the Standard Model neutrinos can be generated through the Seesaw mechanism \cite{Minkowski:1977sc, GellMann:1980vs, Yanagida:1979as, Mohapatra:1979ia, Schechter:1980gr} by including heavy right-handed neutrinos which can easily be implemented in the MSSM.

While collider searches for new physics have remained unsuccessful so far, additional information can be obtained from examining precision observables involving flavour transitions. For example, in the hadronic sector, the branching ratios of rare decays such as $b \to s \gamma$ are sensitive to new physics contributions, especially if they involve non-minimal flavour violation (NMFV), i.e.\ sources of flavour violation beyond the Cabbibo Kobayashi Maskawa (CKM) matrix \cite{Cabibbo1963, Kobayashi1973}. The same holds for the lepton sector \cite{Maki1962, Pontecorvo1967}, where NMFV contributions can induce the branching ratios like $\mu \to e \gamma$ and $\mu \to 3e$, or $\mu-e$ conversion rates in nuclei. Flavour precision observables therefore provide an interesting handle towards the new physics spectrum, and in particular towards the underlying flavour structure. 

Extensive studies have shown that the MSSM parameter space can accomodate NMFV in both the squark and the slepton sectors despite the numerous experimental and theoretical constraints \cite{Arana-Catania:2014ooa, Kowalska:2014opa, DeCausmaecker:2015yca}. In addition, NMFV may lead to specific collider signatures \cite{Bartl:2005yy, Bozzi:2007me, Bartl:2007ua, delAguila:2008iz, Hurth:2009ke, Bartl:2009au, Bruhnke:2010rh, Bartl:2010du, Bartl:2011wq, Bartl:2012tx, Bartl:2014bka}, weaken the current mass limits derived from the non-observation of superpartners \cite{Blanke:2015ulx, Brooijmans:2018xbu, Chakraborty:2018rpn}, and, although to some lesser extend, affect the dark matter phenomenology \cite{Barger:2009gc, Choudhury2011, Herrmann:2011xe, Agrawal:2014aoa}. 

A particularly interesting feature of the MSSM and related models is that they can be successfully embedded into Grand Unified Theories (GUTs). Such a framework allows a unification of the gauge couplings at a scale of about $10^{16}$~GeV with better precision than the Standard Model alone \cite{Amaldi:1991cn}. In the same spirit, the soft breaking parameters related to squarks and sleptons stem from a common origin. In the most simple realizations this allows for the reduction of the number of parameters of the model. Starting from the imposed values at the GUT scale, the phenomenological aspects are obtained through renormalization group running to the TeV scale, where the physical masses and related observables are computed.

The two aforementioned aspects can be addressed by considering SUSY-GUTs including flavour symmetries, such as $SU(5) \times A_4$ \cite{Belyaev:2018vkl} or $SU(5) \times S_4$ \cite{Dimou:2015cmw, Dimou:2015yng}, to cite only two examples. In such a situation, the flavour structure of the theory is defined at the GUT scale by the imposed symmetry. Renormalization group running then translates the GUT-scale structure into the observable mass spectrum at the TeV scale. The TeV-scale phenomenology thus inherits a footprint of the imposed flavour structure at the GUT scale.

In a previous study \cite{Bernigaud:2018qky}, some of the authors have explored the phenomenology of NMFV within a $SU(5) \times A_4$ implementation of the MSSM suggested first in Ref.\ \cite{Belyaev:2018vkl}. Based on the variation of the NMFV parameters around a MFV reference scenario taken from Ref.\ \cite{Belyaev:2018vkl}, it has been shown that these parameters need to be varied simultaneously in order to cover all phenomenological aspects, in particular since cancellations between different contributions may occur. Moreover, it has been demonstrated that a model may feature a reasonable amount of flavour violation while satisfying the stringent constraints of rare decays such as $b \to s \gamma$ or $\mu \to e \gamma$ while sufficient lepton flavour violation may also address baryon asymmetry in the universe \cite{Mukaida:2021sgv}. It is therefore interesting to pursue the study of GUT implementations of flavour violation, e.g., via flavour symmetries, in the context of low-energy and precision constraints as well as TeV-scale phenomenology.

In the present paper, we shall focus on the case of $SU(5)$ unification combined with an $S_4$ flavour symmetry, as suggested by one of the authors in Refs.\ \cite{Dimou:2015cmw, Dimou:2015yng}. In a similar way as in Ref.\ \cite{Bernigaud:2018qky}, we will explore in detail the TeV-scale aspects of this model, including observables related to flavour violation and dark matter phenomenology. More precisely, in this study, we aim at a complete exploration of the associated parameters, i.e.\ including a variation of all relevant parameters at the GUT scale. For the sake of an efficient exploration, we make use of the Markov Chain Monte Carlo technique \cite{Metropolis1953, Hastings1970, Markov1971}.

The paper is organised as follows. In Section\ \ref{sec:model}, we review the assumed model. Section\ \ref{sec:method} is dedicated to the discussion of the Markov Chain Monte Carlo algorithm that we employ to efficiently explore the model parameter space. Results are then presented in Section\ \ref{sec:results}. Our conclusions are given in Section\ \ref{sec:conc}.

\section{The Model}
\label{sec:model}

\subsection{Fields and symmetries}

The model developed in Refs.\ \cite{Hagedorn:2010th, Hagedorn:2012ut, Dimou:2015yng, Dimou:2015cmw} is based on the grand unifying group $SU(5)$ combined with an $S_4$ family symmetry, and supplemented by a $U(1)$ symmetry. 
The left-handed quarks and leptons are unified into the representations $\mathbf{\bar{5}}$, $\mathbf{10}$ and $\mathbf{1}$ of $SU(5)$ according to,
\begin{align}
	F_{\alpha} \sim \overline{\bf 5} \sim  \left( \begin{array}{c} d_r^c \\ d_b^c \\ d_g^c \\ e^- \\ -\nu \end{array} \right)_{\!\!\alpha} \,, 
	\qquad
	T_{\alpha} \sim {\bf 10} \sim  \left( \begin{array}{ccccc} 0 & u_g^c & -u_b^c & u_r & d_r \\ . & 0 & u_r^c & u_b & d_b \\ 
		. & . & 0 & u_g & d_g \\ . & . & . & 0 & e^c \\ . & . & . & . & 0 \end{array}\right)_{\!\!\alpha} \,, \qquad
		N_{\alpha} \sim {\bf 1} \sim \nu^c_{\alpha}
	\label{Eqn:SU5_reps}
\end{align}
where the superscript $c$ stands for $CP$-conjugated fields (which would be right-handed without the $c$ operation), and $\alpha=1, 2, 3$ is a family index. The three families are controlled by a family symmetry $S_4$, with $F$ and $N$ each forming a triplet and the first two families of $T$ forming a doublet, while the third family $T_3$ (containing the top quark) is a singlet, as summarised in Table \ref{tab:field_content}.
The choice of the third family $T_3$ being a singlet, permits a renormalisable top quark Yukawa coupling to the singlet Higgs discussed below.

The $S_4$ singlet Higgs fields $H_{\bf{5}}, \ H_{\bf{\bar{5}}}$ and $H_{\bf{\bar{45}}}$, each contain a doublet $SU(2)_L \times U(1)_Y$ representation that eventually form the standard up ($H_u$) and down ($H_d$) Higgses of the MSSM (where the $H_d$
emerges as a linear combination of doublets from the $H_{\bf{\bar{5}}}$ and $H_{\bf{\bar{45}}}$) ~\cite{Chung:2003fi}.\footnote{As $H_{\bf{\bar  5}}$ and $H_{\bf{\bar{45}}}$ transform differently under $U(1)$, it is clear that the mechanism which spawns the low energy Higgs doublet $H_d$ must necessarily break $U(1)$. Although the discussion of any details of the $SU(5)$ GUT symmetry breaking (which, e.g., could even have an extra dimensional origin) are beyond the scope of our paper, we remark that a mixing of $H_{\bf{\bar 5}}$ and  $H_{\bf{\bar{45}}}$ could be induced by  introducing the pair $H^\pm_{\bf{24}}$ with $U(1)$ charges $\pm 1$ in addition to the standard $SU(5)$ breaking Higgs $H^0_{\bf{24}}$.} The VEVs of the two neutral Higgs fields are
\begin{equation}
v_u = \frac{v}{\sqrt{1+t_\beta^2}} \, t_\beta, ~~~~~~v_d=\frac{v}{\sqrt{1+t_\beta^2}},
\end{equation}
where $t_\beta\equiv \tan(\beta)=\frac{v_u}{v_d}$ and $v = \sqrt{v_u^2+v_d^2} \approx 246$ GeV.

\begin{table}
\centering
\scalebox{0.93}{
	\begin{tabular}{c||c|c|c|c||c|c|c||c|c|c|c|c|c|c|c|c|}
		Field & $T_3$ & $T$ & $F$ & $N$ & $H_5$ & $H_{\overline{5}}$ & $H_{\overline{45}}$ & $\Phi_2^u$ & $\widetilde{\Phi}_2^u$ & $\Phi_{3}^{d}$ & $\widetilde{\Phi}_3^d$ & $\Phi_2^d$ & $\Phi^{\nu}_{3'}$ & $\Phi^{\nu}_{2}$ & $\Phi^{\nu}_{1}$ & $\eta$\\
		\hline
		$SU(5)$ & $\mathbf{10}$ & $\mathbf{10}$ & $\mathbf{\overline{5}}$ & $\mathbf{1}$ & $\mathbf{5}$ & $\mathbf{\overline{5}}$ & $\mathbf{\overline{45}}$ & $\mathbf{1}$ & $\mathbf{1}$ & $\mathbf{1}$ & $\mathbf{1}$ & $\mathbf{1}$ & $\mathbf{1}$ & $\mathbf{1}$ & $\mathbf{1}$ & $\mathbf{1}$ \\
		\hline
		$S_4$ & $\mathbf{1}$ & $\mathbf{2}$ & $\mathbf{3}$ & $\mathbf{3}$ & $\mathbf{1}$ & $\mathbf{1}$ & $\mathbf{1}$ & $\mathbf{2}$ & $\mathbf{2}$ & $\mathbf{3}$ & $\mathbf{3}$ & $\mathbf{2}$ & $\mathbf{3'}$ & $\mathbf{2}$ & $\mathbf{1}$ & $\mathbf{1'}$ \\
		\hline
		$U(1)$ & 0 & 5 & 4 & -4 & 0 & 0 & 1 & -10 & 0 & -4 & -11 & 1 & 8 & 8 & 8 & 7 \\
		\hline
	\end{tabular}
	}
	\caption{Field content of the model and associated charges and representations.}
	\label{tab:field_content}
\end{table}

Just below the $SU(5)$ breaking scale to the usual SM gauge group, the flavour symmetry is broken by the VEVs of some new fields: the flavons, $\Phi^f_{\rho}$, which are
labelled by the corresponding $S_4$ representation~$\rho$ as well as the fermion
sector~$f$ to which they couple at leading order (LO). 
Two flavons, $\Phi^u_2$ and $\tilde{\Phi}^u_2$, generate the LO up-type quark
mass matrix. Three flavon multiplets, $\Phi^d_3$, $\tilde{\Phi}^d_3$ and
$\Phi^d_2$, are responsible for the down-type quark and charged lepton mass
matrices. Finally, the right-handed neutrino mass matrix is generated from
the flavon multiplets $\Phi^\nu_{3'}$,  $\Phi^\nu_2$ and $\Phi^\nu_1$  as well as
the flavon $\eta$ which is responsible for breaking the tri-bimaximal pattern
of the neutrino mass matrix to a trimaximal one at subleading
order. An additional $U(1)$ symmetry must be introduced in order
to control the coupling of the flavon fields to the matter fields in a way
which avoids significant perturbations of the flavour structure by
higher-dimensional operators. 

\subsection{Flavon alignments}

The vacuum alignment of the flavon fields
is achieved by coupling them to a set of so-called driving fields and requiring
the $F$-terms of the latter to vanish. These driving fields, whose
transformation properties under the family symmetry are discussed in Refs.\ \cite{Hagedorn:2010th, Hagedorn:2012ut, Dimou:2015yng, Dimou:2015cmw}, are SM gauge singlets and carry a charge of $+2$
under a continuous $R$-symmetry. The flavons and the GUT Higgs fields are
uncharged under this $U(1)_R$, whereas the supermultiplets containing the SM
fermions (or right-handed neutrinos) have charge $+1$. As the superpotential
must have a $U(1)_R$ charge of $+2$, the driving fields can only appear
linearly and cannot have any direct interactions with the SM fermions or
the right-handed neutrinos. 

Using the driving fields, the flavour superpotential may be constructed,
resulting in the following vacuum alignments 
(for details see Refs.\ \cite{Hagedorn:2010th, Hagedorn:2012ut}),
\begin{eqnarray}
\label{LeadingFlavonVevs1}
\frac{\langle \Phi^u_2 \rangle}{M} &=& \left(
\begin{array}{ccc}
0\\[-0.5mm]
1
\end{array}
\right)\phi^u_2\,\lambda^4,
\qquad
 \frac{\langle \tilde{\Phi}^u_2 \rangle}{M} = \left(
\begin{array}{ccc}
0\\[-0.5mm]
1
\end{array}
\right)\tilde{\phi}^u_2\,\lambda^4,\\[-1mm]
\label{LeadingFlavonVevs2}
 \frac{\langle \Phi^d_3 \rangle}{M}&=&\left(
\begin{array}{ccc}
0\\[-0.5mm]
1\\[-0.5mm]
0
\end{array}
\right)\phi^d_3\,\lambda^2,
\qquad
\frac{\langle \tilde{\Phi}^d_3 \rangle}{M}=\left(
\begin{array}{ccc}
0\\[-0.5mm]
-1\\[-0.5mm]
1
\end{array}
\right) \tilde{\phi}^d_3\,\lambda^3, 
\qquad
\frac{\langle \Phi^d_2 \rangle}{M}=\left(
\begin{array}{ccc}
1\\[-0.5mm]
0
\end{array}
\right)\phi^d_2\,\lambda\ ,\\[-1mm]
\label{LeadingFlavonVevs3}
\frac{\langle \Phi^\nu_{3'} \rangle}{M}&=&\left(
\begin{array}{ccc}
1\\[-0.5mm]
1\\[-0.5mm]
1
\end{array}
\right)\phi^\nu_{3'}\,\lambda^4,
\qquad
\frac{\langle \Phi^\nu_2 \rangle}{M}=\left(
\begin{array}{ccc}
1\\[-0.5mm]
1
\end{array}
\right)\phi^\nu_2\,\lambda^4,
~\quad
 \frac{\langle \Phi^\nu_1 \rangle}{M}= \phi^{\nu}_1\,\lambda^4 ,
~\quad
 \frac{\langle \eta \rangle}{M} = \phi^\eta\,\lambda^4,~~~~~~
\end{eqnarray}\\[-3.5mm]
where $\lambda = 0.22$ is approximately equal to the Wolfenstein
parameter \cite{Wolfenstein:1983yz} and the $\phi$'s are dimensionless order 
one parameters. Imposing $CP$-symmetry of the underlying theory \cite{Luhn:2013vna}, all coupling constants can be taken real, so that $CP$ is broken
spontaneously by generally complex values for the $\phi$s. The remaining phases are independent and there is no residual $CP$-symmetry. $M$~denotes a generic messenger scale which is common to all the non-renormalisable effective operators and assumed to be around the scale of grand unification. 

\subsection{Yukawa matrices}
\label{ChargedFermionSector}

Because of the non-trivial structure of the Kähler potential, non-canonical kinetic terms are generated. For a proper analysis of the flavour structure, one needs to perform a canonical normalisation (CN) operation, swapping the misalignment of the kinetic terms to the superpotential. Therefore, in the model proposed in Refs.\  \cite{Hagedorn:2010th, Hagedorn:2012ut, Dimou:2015yng, Dimou:2015cmw}, 
contributions to the flavour texture from both the superpotential and the Kähler potential are taken into account. In this subsection, we shall begin by ignoring such corrections, 
and also only consider the leading order Yukawa operators,
in order to clearly illustrate the origin of the flavour structure in the model. However, all such corrections are taken into account
in the phenomenological treatment of the Yukawa matrices in the following subsection.
We remark that the model is highly predictive, as the parameters entering the flavour structure are expected to be of $O(1)$ but the overall flavour texture is provided as a function of the expansion parameter $\lambda = 0.22$. 

\subsubsection{Up-type quarks}

The Yukawa matrix of the up-type quarks can be constructed by considering all
the possible combinations of a product of flavons with $TTH_5$ for the
upper-left $2\times2$ block, with $TT_3H_5$ for the ($i3$) elements, and
with $T_3T_3H_5$ for the (33) element. The most important operators which generate a
contribution to the Yukawa matrix of order up to and including $\lambda^8$ are
\begin{equation}
y_tT_3 T_3 H_5+\frac{1}{M}y^u_1TT\Phi^u_2 H_5+\frac{1}{M^2}y^u_2TT\Phi^u_2\tilde{\Phi}^u_2 H_5 \,,
\label{YuOperators}
\end{equation}%
where the parameters $y_t$ and $y^u_{i}$ are real order one coefficients.
Inserting the flavon VEVs and expanding the $S_4$ contractions of
Eq.~\eqref{YuOperators}, with $TT$ and $\Phi^u_2\tilde{\Phi}^u_2$ each combined into a doublet using the Clebsch-Gordan coefficients \cite{Hagedorn:2010th,Hagedorn:2012ut}, yields the up-type Yukawa matrix at the GUT scale
\begin{eqnarray}
\mathcal Y^u_{\text{GUT}}&\approx&\left(
\begin{array}{ccc}
 y_u e^{i\theta^y_u}\lambda ^8 & 0 & 0 \\
 0 &y_c e^{i\theta^y_c} \lambda ^4&  0\\
 0 &  0 & y_t 
\end{array}
\right) \ ,\label{Yu}
\end{eqnarray}%
where the relation to the flavon VEVs,
see Eqs.\ \eqref{LeadingFlavonVevs1} -- \eqref{LeadingFlavonVevs3}, is given by 
\begin{equation}
 y_u e^{i\theta^y_u}=y^u_2\phi^u_2\tilde{\phi}^u_2
\,,\qquad
y_c \,e^{i\theta^y_c}=y^u_1\phi^u_2 \,.
\label{dexerw1}
\end{equation}

\subsubsection{Down-type quarks and charged leptons}

The Yukawa matrices of the down-type quarks and the charged leptons can be
deduced from the leading superpotential operators
\begin{equation}
y^d_1\frac{1}{M}FT_3\Phi^d_3H_{\bar{5}}+
y^d_2\frac{1}{M^2}(F\tilde{\Phi}^d_3)_{\mathbf 1}(T\Phi^d_2)_{\mathbf 1}H_{\bar{45}}
+y^d_3\frac{1}{M^3}(F(\Phi^d_2)^2)_{\mathbf 3}(T\tilde{\Phi}^d_3)_{\mathbf 3}H_{\bar{5}} \,,
\label{YdOperators}
\end{equation}
where the $y^d_i$ are real order one coefficients. For the operators
proportional to $y^d_2$ and $y^d_3$, specific $S_4$ contractions 
indicated by $(\cdots)_{\mathbf 1}$ and $(\cdots)_{\mathbf 3}$ have been chosen 
(justified by messenger arguments)
such that the Gatto-Sartori-Tonin (GST) \cite{Gatto:1968ss} and Georgi-Jarlskog (GJ) \cite{Georgi:1979df} relations are satisfied. 
Separating the contributions of $H_{\bf{\bar{5}}}$  and
$H_{\bf{\bar{45}}}$, the $S_4$ contractions give rise to 
\begin{equation}
\mathcal Y_{\bf{\bar{5}}}\approx\left(
 \begin{array}{ccc}
 0 &\tilde{x}_2 e^{i\theta^{\tilde{x}}_2} \lambda^5&-\tilde{x}_2 e^{i\theta^{\tilde{x}}_2} \lambda^5\\
 -\tilde{x}_2 e^{i\theta^{\tilde{x}}_2} \lambda^5& 0 &\tilde{x}_2 e^{i\theta^{\tilde{x}}_2} \lambda^5 \\
0&0&y_be^{i\theta^y_b} \lambda^2
\end{array}
\right)  ,~\quad
\mathcal Y_{\bf{\bar{45}}} \approx \left(
\begin{array}{ccc}
 0&0&0\\
0& y_s e^{i\theta^y_s}\lambda^4 &-y_s e^{i\theta^y_s}\lambda^4 \\
 0 & 0&0
\end{array}
\right)  .~ \label{Ydb}
\end{equation}
The parameters in these expressions are related to the flavon VEVs
as defined in Eqs.\ \eqref{LeadingFlavonVevs1}--\eqref{LeadingFlavonVevs3} via
\begin{equation}
y_be^{i\theta^y_b}=y^d_1\phi^d_3\, ,~\quad 
y_se^{i\theta^y_s}=y^d_2\phi^d_2\tilde{\phi}^d_3,\, , ~\quad
\tilde{x}_2e^{i\theta^{\tilde{x}}_2}=y^d_3(\phi^d_2)^2\tilde{\phi}^d_3 \,.
\label{ax}
\end{equation}

The Yukawa matrices of the down-type quarks and the charged leptons 
are linear combinations of the two structures in
Eq.\ (\ref{Ydb}). Following the construction proposed by Georgi
and Jarlskog, we have
$\mathcal Y^d_{\text{GUT}}=\mathcal Y_{\bar{5}}+\mathcal Y_{\bar{45}}, \ \ \ \ 
\mathcal Y^e_{\text{GUT}}=(\mathcal Y_{\bar{5}}-3\mathcal Y_{\bar{45}})^T$.
CKM mixing is
dominated by the diagonalisation of the down-type quark Yukawa matrix. 
Note that in some models it is possible to go beyond the simple case $m_b=m_{\tau}$ at the GUT scale, by including larger Higgs representations \cite{Antusch:2009gu, Antusch:2013rxa}.

\subsubsection{Neutrinos}

The neutrino masses originate from a standard Supersymmetric Type I Seesaw mechanism, where the heavy right-handed fields, N, are turning the tiny observed neutrino effective Yukawa couplings into natural parameters. The Lagrangian for the neutrino sector is therefore given by
\begin{align}
\mathcal{L}_{\nu} \supset ( \mathcal Y^\nu)_{ij}\overline{L}_iH_uN_j + (\mathcal M_R)_{ij}N_iN_j\,,
\label{eqn:seesaw_Lagrangian}
\end{align}
where $\mathcal Y_\nu$ is the Dirac Yukawa coupling and $\mathcal M_R$ is the right-handed Majorana mass matrix.

The Dirac coupling of the right-handed neutrinos $N$ to the left-handed SM neutrinos is dominated by the
superpotential term
\begin{equation}
y_D FNH_5
\label{YnuOperators}
\end{equation}%
where $y_D$ is a real order one parameter. The corresponding Yukawa matrix is determined as
\begin{eqnarray}
 \mathcal Y^\nu%
&\approx&\left(
\begin{array}{ccc}
y_D&0&0\\
0&0&y_D\\
0&y_D&0
\end{array}
\right) \ , \label{Ynu}
\end{eqnarray}%

The mass matrix of the right-handed neutrinos is obtained from the
superpotential terms
\begin{equation}
w_{1,2,3}NN\Phi^\nu_{1,2,3'}+w_4\frac{1}{M}NN\Phi^d_2\eta \ , 
\label{YNOperators}
\end{equation}%
where  $w_i$ denote real order one coefficients.  This results in a
right-handed Majorana neutrino mass matrix $\mathcal M_R$ of the form
\begin{eqnarray}
\frac{\mathcal M_R}{M}&\approx&\left(
\begin{array}{ccc}
A+2C&~~~B-C &~~~B-C\\
B-C&~~~B+2C &~~~A-C\\
B-C&~~~A-C&~~~B+2C
\end{array}
\right)e^{i\theta_A}\lambda^4+
\left(\begin{array}{ccc}
0&~0&~D\\
0&~D&~0\\
D&~0&~0
\end{array}
\right)e^{i\theta_D}\lambda^5 \, , ~~~~~~~ \label{MR}
\end{eqnarray}%
with 
\begin{equation}
Ae^{i\theta_{A}}\!=w_1\phi^\nu_1\, , \,\quad
Be^{i\theta_A}\!=w_2\phi^\nu_2\, , \,\quad
Ce^{i\theta_A}\!=w_3\phi^\nu_{3'}\, , \,\quad
De^{i\theta_D}\!=w_4\,\eta\,\phi^d_2\,. \label{ABCD}
\end{equation}

The first matrix of Eq.~\eqref{MR} arises from terms involving only
$\Phi^\nu_{1,2,3'}$. As their VEVs respect the tri-bimaximal (TB) Klein symmetry $Z_2^S\times Z_2^U\subset S_4$, this part is of TB form. The second matrix of
Eq.\ \eqref{MR}, proportional to $D$, is due to the operator
$w_4\frac{1}{M}NN\Phi^d_2\eta$ which breaks the $Z_2^U$ at a relative
order of $\lambda$, while preserving the $Z_2^S$. The resulting trimaximal TM$_2$ structure can
accommodate the sizable value of the reactor neutrino mixing angle
$\theta^l_{13}$.

It is instructive to show the effective light neutrino mass matrix which arises via the type~I seesaw mechanism, and has the form
\begin{eqnarray}\label{eq:efflightnu}
 m^{\text{eff}}_{\nu}&\approx&\frac{y_D^2 v_u^2}{\lambda^4 M}\left[ \! \left(
\begin{array}{ccc}
b^\nu +c^\nu-a^\nu &a^\nu&a^\nu\\
a^\nu&b^\nu &c^\nu \\
a^\nu &c^\nu&b^\nu
\end{array}
\right)e^{-i\theta_A}+\left(
\begin{array}{ccc}
0&0&d^\nu\\
0&d^\nu&0\\
d^\nu &0&0
\end{array}
\right)\lambda\,e^{i(\theta_D-2\theta_A)} \right]  ,~~~~~~
\end{eqnarray}%
with $a^\nu$, $b^\nu$, $c^\nu$ and $d^\nu$ being functions of the real parameters $A$, $B$, $C$ and $D$. The deviation from tri-bimaximal neutrino mixing is controlled by $d^\nu \propto D$. Due to the three independent input parameters ($w_1\propto A$, $w_2\propto B$, $w_3\propto C$), any neutrino mass spectrum can be accommodated in this model. However, in our numerical analysis, we shall restrict our scans to the case of a normal neutrino mass ordering, which is preferred by the latest global fits of neutrino oscillation data. Note that this expression, while generally providing a good estimation of the effective neutrino mass matrix, is only an approximation valid at order $D\times\lambda$ and therefore, it does not strictly hold when considering a potential $O(10)$ $D$-parameter. This is why this expression is more for illustrative purposes while we performed a rigorous treatment of the full seesaw mechanism in the numerical analysis. 

\subsection{Phenomenological Yukawa couplings at the GUT scale}

The true model predictions at the high scale differ from those shown previously, since they 
also involve other higher order corrections to the Yukawa terms, and one must also include the effects of canonical normalisation (CN) leading to the matrices in Ref.\ \cite{Dimou:2015yng}. For simplicity, while keeping the phenomenology indistinguishable from the constructed model, we allow for minor approximations, and here we summarise the form of the Yukawa matrices that we actually assume at the GUT scale.

Concerning the up-type quark Yukawa matrix, we shall continue to take it to be diagonal as the off-diagonal entries are much more $\lambda$ suppressed than the diagonal ones. We may also absorb the phases into a redefinition of the fields.

Since the CKM matrix is controlled by the down-type quark Yukawa matrix, we shall include some of the higher-order terms and some of the effects of CN, in order to obtain a perfect fit to quark data. Therefore there are some corrections to the GST relations.

The charged lepton Yukawa matrix is, like in any standard $SU(5)$ model, closely related to the down-quark Yukawa matrix as per $Y_{\ell} \simeq Y_{d}^T$, together with a modified via the GJ mechanism through the incorporation of $\mathbf{\bar{45}}$ and $\mathbf{\bar{5}}$ Higgs representations in order to generate a more reasonable relation between $m_s$ and $m_\mu$. 

The explicit Yukawa matrices we will use for the charged fermionic sector are therefore provided by the following expressions:
\begin{align}
	\begin{split}
		&Y_u =
		\begin{pmatrix}
		y_{u}\lambda^8 & 0 & 0\\
		0 & y_{c}\lambda^4 & 0\\
		0 & 0 & y_{t}
		\end{pmatrix},
		\\
		&Y_d =
		\begin{pmatrix}
		z_{1}^{d} \lambda^8\, e^{-i\delta} & x_{2}\lambda^5 & -x_{2}\lambda^{5} \, e^{i\delta}\\
		-x_{2}\lambda^{5} & y_{s}\lambda^4\, e^{-i\theta_{2}^d} & -y_{s}\lambda^4\,e^{2i(\theta_{2}^d+\theta_{3}^d)} + x_2 \lambda^5 e^{3i(\theta_{2}^d + \theta_{3}^d)}\\
		(z_{3}^{d}\, e^{-i\theta_{2}^d}-\frac{1}{2}K_{3}y_{b}\, e^{-i\delta})\lambda^{6} & (z_{2}^{d}\, e^{-i\theta_{2}^d}-\frac{1}{2}K_{3}y_{b}\, e^{-i\delta})\lambda^{6} & y_{b}\lambda^2
		\end{pmatrix},\\[10pt]
		&Y_{\ell} =
		\begin{pmatrix}
		-3 {z_{1}^{d}} \lambda^8\, e^{-i\delta} & - x_{2}\lambda^5 & (z_{3}^{d}\, e^{-i\theta_{2}^d}-\frac{1}{2}K_{3}y_{b}\, e^{-i\delta})\lambda^{6}\\
		x_{2}\lambda^{5} & -3 y_{s}\lambda^4\, e^{-i\theta_{2}^d} & (z_{2}^{d}\, e^{-i\theta_{2}^d}-\frac{1}{2}K_{3}y_{b}\, e^{-i\delta})\lambda^{6}\\
		-x_{2}\lambda^{5} \,e^{i\delta} &3y_{s}\lambda^4\,e^{2i(\theta_{2}^d+\theta_{3}^d)}+ x_2 \lambda^5 e^{3i(\theta_{2}^d + \theta_{3}^d)}  & y_{b}\lambda^2
		\end{pmatrix} \,,
	\end{split}
	\label{eqn:model_Yukawa_couplings}
\end{align}
where $\delta = 2\theta_{2}^d + 3\theta_{3}^d$, and $\theta_{\rho}^{d}$ ($\rho=2,3$) corresponds to the phase of a $\rho$-representation flavon in the original model. Note that our analysis, including the soft masses discussed below, relies only on the two phases $\theta_{2}^d$ and $\theta_{3}^d$. These Yukawa matrices are obtained using Eq.\ \eqref{Ydb}. However, one needs to perform a canonical normalisation of the kinetic terms. This procedure has been taken care of in Ref.\ \cite{Dimou:2015yng}. Therefore, our phenomenological Yukawa couplings involve more parameters than the ones exposed in Eq.\ \eqref{Ydb}. The up-type quark Yukawa has been approximated with respect to Ref.\ \cite{Dimou:2015yng}. 

In the neutrino sector, the effects of CN are negligible, and we therefore take these matrices to have the same form as given previously,
\begin{equation}
\begin{aligned}
\frac{M_R}{M_{GUT}} =
\begin{pmatrix}
A+2C& B-C & B-C\\
B-C & B+2C & A-C\\
B-C & A-C & B+2C
\end{pmatrix}\lambda^4\,e^{-2i\theta_{3}^d} + \begin{pmatrix}
0 & 0 & D \\
0 & D & 0 \\
D & 0 & 0
\end{pmatrix}\lambda^5\, e^{i(4\theta_{2}^d - \theta_{3}^d)}\,.
\end{aligned}
\label{eqn:neutrino_couplings_UV}
\end{equation}
The Dirac neutrino coupling, neglecting the $O(\lambda^4)$ terms, compared to the original paper \cite{Dimou:2015yng}, that is also of the form given in the previous subsection,
\begin{equation}
Y_\nu =
\begin{pmatrix}
y_{D} & 0 & 0\\
0 & 0 &y_{D}\\
0 & y_{D} & 0
\end{pmatrix} \,.
\label{Dirac}
\end{equation}

\subsection{SUSY breaking terms}

We now consider the SUSY breaking sector of the low energy scale MSSM generated after integrating out the heavy degrees of freedom. In the context of the standard phenomenological $R$-parity conserving MSSM, the soft Lagrangian is parametrised as
\begin{align}
	\begin{split}
		\mathcal{L}^{\rm MSSM}_{\rm soft} = 
		& - \frac{1}{2} \big( M_1\widetilde{B}\widetilde{B}+M_2\widetilde{W}\widetilde{W} 
		+ M_3\widetilde{g}\widetilde{g} + \rm{h.c.} \big) \\[1ex]
		& - M_Q^2\widetilde{Q}^{\dagger}\widetilde{Q} - M_L^2\widetilde{L}^{\dagger}\widetilde{L}
		- M_U^2\widetilde{U}^*\widetilde{U} - M_D^2\widetilde{D}^*\widetilde{D} 
		- M_E^2\widetilde{E}^*\widetilde{E} - M_N^2\widetilde{N}^*\widetilde{N} \\[1ex]
		& - \big( A_U \widetilde{U}^*H_u\widetilde{Q} + A_D\widetilde{D}^*H_d\widetilde{Q}
		+ A_E \widetilde{E}^*H_d\widetilde{L} + A_\nu \widetilde{N}^*H_u\widetilde{L} + \rm{h.c.} \big) \\[1ex]
		& - m_{H_u}^2 H_u^*H_u - m_{H_d}^2 H_d^*H_d - \big( b H_u H_d+{\rm h.c.} \big) \,.
	\end{split}
	\label{eqn:SUSY_breaking_Lagrangian}
\end{align}
where $\Tilde{\phi}$ denotes the generic superpartner of a generic SM particle $\phi$. Assuming that the SUSY breaking is controlled by some hidden sector mediated by a superfield $X$, the soft parameters described in \eqref{eqn:SUSY_breaking_Lagrangian} are generated when $X$ develops a VEV in its F-term at the SUSY breaking scale. Furthermore, we consider that the SUSY breaking mechanism is independent of the flavour breaking one. Note that we assume non universal gaugino masses in our analysis. While this is the case in simplistic $SU(5)$ models, many extensions can account for non universal masses, see e.g.\ Ref.\ \cite{King:2007vh}.

The new flavour structure arising from the SUSY breaking sector is also controlled by the flavour symmetry, in a similar fashion as the SM texture is. 
Extracting the results from Ref.\ \cite{Dimou:2015yng}, we first summarise the predictions for the soft trilinear terms $A_f$,
\begin{align}
\begin{split}
	\frac{A_{GUT}^u}{A_0} &\simeq \begin{pmatrix}
		a_u\lambda^8 & 0 & 0\\
		0 & a_c\lambda^4 & \\
		0 & 0 & a_t
	\end{pmatrix} \,,\\[10pt]
	\frac{A_{GUT}^d}{A_0} &\simeq \begin{pmatrix}
		z_1^{d_a}\lambda^8 & \widetilde{x}_2^a\lambda^5 & -\widetilde{x}_2^a\lambda^5\\
		-\widetilde{x}_2^a\lambda^5 & a_s\lambda^4 & -a_s\lambda^4\\
		(z_3^{d_a}-\frac{K_3a_b}{2})\lambda^6 & (z_2^{d_a}-\frac{K_3a_b}{2})\lambda^6 & a_b\lambda^2
	\end{pmatrix} \,,\\[10pt]
	\frac{A_{GUT}^e}{A_0} &\simeq \begin{pmatrix}
		-3a_d\lambda^8 & -\widetilde{x}_2^a\lambda^5 & (z_3^{d_a}-\frac{K_3a_b}{2})\lambda^6 \\
		\widetilde{x}_2^a & -3a_s\lambda^4 & (z_2^{d_a}-\frac{K_3a_b}{2})\lambda^6 \\
		-\widetilde{x}_2^a\lambda^5 & 3a_s\lambda^4 & a_b\lambda^2
	\end{pmatrix} \,,\\[10pt]
	\frac{A_{GUT}^\nu}{A_0} &\simeq \begin{pmatrix}
		a_D & 0 & 0\\
		0 & 0 & a_D\\
		0 & a_D & 0
	\end{pmatrix} \,.
	\end{split}
\end{align}
The trilinear soft couplings exhibit the same structure as the Yukawa terms, except that the $O(1)$ parameters are now different. Note that in our numerical analysis we neglect the phases appearing in the trilinear matrices. We have verified that such phases have a negligible affect on the $CP$-conserving constraints. We further use the same approximations as the ones considered for the Yukawa couplings. 

Similarly, we summarise the results on the soft scalar mass matrices,
\begin{align}
\begin{split}
	M_T^2 &\simeq m_0^2 \begin{pmatrix}
		b_{01} & (b_{2}-b_{01}k_2)\lambda^4 & (b_4-\frac{k_4(b_{01}-b_{02})}{2})e^{-i\theta_{4k}}\lambda^6\\
		\cdot & b_{01} & (b_3-\frac{k_3(b_{01}-b_{02})}{2})e^{-i\theta_{3k}}\lambda^5\\
		\cdot & \cdot & b_{02}
	\end{pmatrix} \,, \\[10pt]
	M_{F(N)}^2 &\simeq m_0^2 \begin{pmatrix}
	B_{0}^{(N)} & (B_3^{(N)}-K_3^{(N)})\lambda^4 & (B_3^{(N)}-K_3^{(N)})\lambda^4 \\
	\cdot & B_{0}^{(N)} & (B_3^{(N)}-K_3^{(N)})\lambda^4 \\
	\cdot & \cdot & B_{0}^{(N)}
	\end{pmatrix} \,,
	\label{Eq:Soft_matrices}
	\end{split}
\end{align}
where $N$ is the right handed sneutrino term, $\theta_{3k}=-5\theta_{2}^d$ and $\theta_{4k}=\theta_3^d-\theta_2^d$. Because of the unification, all the sfermion soft matrices are linked to the soft matrix of the $SU(5)$ representation they belong to.
\section{Data-driven model exploration}
\label{sec:method}

\subsection{Algorithm}

The full analysis of the parameter space relies on a Markov-Chain Monte Carlo technique \cite{Markov1971}, and more specifically the Metropolis-Hastings algorithm \cite{Metropolis1953, Hastings1970}. This technique allows one to perform a sophisticated data-driven exploration of an high-dimension parameter space. The idea behind the algorithm is to estimate the likelihood $\mathcal{L}$ of a given set of parameter values $\Vec{\theta}$ with respect to the set of observables $\Vec{O}$. For simplicity and the rest of the analysis we assume that the observables are not correlated, i.e.
\begin{equation}
    \mathcal{L}(\Vec{\theta},\Vec{O}, \Vec{\sigma}) = \prod_{i} \mathcal{L}_i(\Vec{\theta},O_i, \sigma_i),
    \label{eq:likelihood}
\end{equation}
where $\sigma_i$ is the uncertainty associated to the observable $O_i$.

Successively, random values of the parameters, picked around the previous ones, are evaluated at each iteration. In our implementation, the new proposed parameter value $\theta^{n+1}$ is obtained through a Gaussian jump,
\begin{equation}
    \theta_{i}^{n+1} = \mathcal{G}\left(\theta_{i}^{n},\, \kappa(\theta_{i}^{ \text{max}} - \theta_{i}^{ \text{min}}) \right)\,,
\end{equation}
where $\mathcal{G}\left(a,b\right)$ is a Gaussian distribution centered around $a$ with width $b$, $\kappa$ is a parameter that needs to be tuned empirically for the algorithm and $\theta_{i}^{ \text{max}}$ and $\theta_{i}^{ \text{min}}$ stands for the extrema values of the $\theta_{i}$ considered range.

If $\mathcal{L}^{n+1}(\Vec{\theta}^{n+1}, \Vec{O}, \vec{\sigma}) > \mathcal{L}^{n}(\Vec{\theta}^n, \Vec{O}, \vec{\sigma})$, the point is accepted and the chain continues from this point. Otherwise, the new point is accepted with probability
\begin{equation}
    p = \frac{\mathcal{L}^{n+1}(\Vec{\theta}^{n+1}, \Vec{O}, \Vec{\sigma})}{\mathcal{L}^{n}(\Vec{\theta}^{n}, \Vec{O}, \Vec{\sigma})} \,
\end{equation}
This avoids falling into local minima, and thus allows for a better parameter space exploration. In practice, we randomly choose a number $\mu \in [0,1]$ such that the test succeeds if $\mu < p$. Otherwise, the point is rejected, and we reevaluate the step $n+1$ for another proposal set of parameters deduced from step $n$. Within this framework, the algorithm can move across larger regions while still converging to highest likelihood regions.

\begin{figure}
    \centering
    \scalebox{0.8}{
    \begin{tikzpicture}[node distance=2cm,
    every node/.style={fill=white, font=\sffamily}, align=center]
    \node (start)  [activityStarts] {Point $n$: \ \  $\Vec{\theta}^{n}$};

    \node (proposal)  [activityStarts, below of=start, yshift=-2cm] {Proposal $n+1$: \ \  $\Vec{\theta}^{n+1}$};

    \node (likelihood)  [activityStarts, below of=proposal, yshift=-2cm] {Likelihood: \ \  $\mathcal{L}^{n+1}(\Vec{\theta}^{n+1},\, \vec{O},\, \vec{\sigma})$};

    \node (Test)  [activityStarts, below of=likelihood, yshift=-2cm] {Test: \ \  $\mu <\frac{\mathcal{L}^{n+1}(\Vec{\theta}^{n+1},\, \vec{O},\, \vec{\sigma})}{\mathcal{L}^n(\Vec{\theta}^{n},\, \vec{O}, \,\vec{\sigma})}$};
    \node (fail)  [activityStarts, left of=Test, xshift=-3.8cm] {Fail:\ \  restart at $n$};
    \node (success)  [activityStarts, right of=Test, xshift=3.8cm] {Success:\ \  $n\rightarrow n+1$};
     
    \draw[->, line width=0.3mm]      (start) -- node[text width=4cm]
                                   { Jump\\ $\mathcal{G}\left(\theta_{i}^{n},\, \kappa\theta_{i}^{n} \right)$} (proposal);
    \draw[->, line width=0.3mm]      (likelihood) -- node {$\mu \in [0,1]$} (Test);
    \draw[->, line width=0.3mm]      (proposal) -- node {$\vec{O}(\vec{\theta}^{n+1})$} (likelihood);
    \draw[red, line width=0.3mm, ->]      (Test) -- (fail);
    \draw[red, ->, line width=0.3mm]      (fail) |- (start);
    \draw[blue, ->, line width=0.3mm]      (Test) -- (success);
    \draw[blue, ->, line width=0.3mm]      (success) |- (start);

    \end{tikzpicture}
    }
    \caption{Illustration of the MCMC algorithm utilisation.}
    \label{fig:MCMC}
\end{figure}
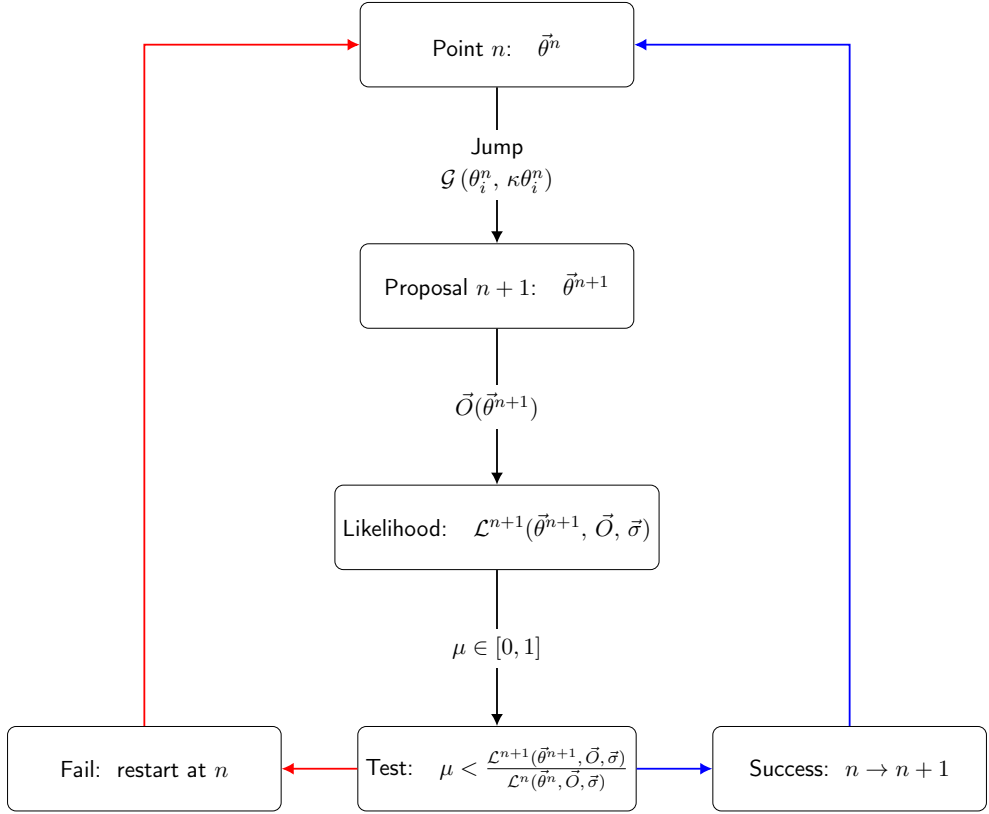

In high-dimensional parameter space, the quality of the exploration relies more on the total of chain numbers than the length of the chain themselves. Indeed, different starting points (chosen randomly) can lead to different likelihood maximums. A summary of the algorithm is given in Figure \ref{fig:MCMC}.

\subsection{Constraints, tools and setup}

We now develop on the numerical tools and constraints employed in our analysis. As the model is defined at the GUT scale, we first perform the evolution of the renormalisation group equations (RGEs)  to the low scale, to derive low energy observables. For this purpose we employed the {\tt SARAH v4.14.1} Mathematica package \cite{Staub:2008uz, Staub:2013tta, Staub:2012pb, Staub:2010jh, Staub:2009bi} in order to generate a type I Seesaw GUT MSSM model based on {\tt SPheno v4.0.4} \cite{Porod:2003um, Porod:2011nf}. Right handed Majorana neutrinos, which typically live near the GUT scale are therefore consistently integrated out at their mass scale. Furthermore, the {\tt Flavour kit} \cite{Porod:2014xia} available within {\tt SARAH} / {\tt SPheno} computes a wide range of flavour observables, simplifying our framework as both one-loop masses, two-loop Higgs mass \cite{Goodsell:2014bna, Goodsell:2015ira, Goodsell:2016udb, Braathen:2017izn} and flavour observables are evaluated within a single executable.

However, modifications of this model have been realised: In the usual {\tt SPheno} instances, SM fermion masses are enforced to match the experimental data by several runs up and down between the GUT scale and the low scale, rendering our model predictions impossible to estimate. To overcome this, we have removed the SM fermions from this iterative convergence process while keeping the massive gauge bosons. To consistently implement such restrictions, attention must be payed concerning several features. An extended discussion regarding our modified {\tt SPheno} version can be found in Appendix \ref{app:spheno}. 

We have also decided to include dark matter constraints in our analysis. Restricting ourselves to neutralino dark matter, we imposed a step dark matter candidate likelihood (1 if the LSP is the lightest neutralino, 0 otherwise). In order to derive relic density and direct detection constraints, we have used {\tt micrOMEGAs v5.2}, which accepts the spectrum files generated by {\tt SPheno} through the SUSY Les Houches Accord (SLHA) \cite{Skands:2003cj, Allanach:2008qq} interface.

It should be noted that {\tt SARAH} generated {\tt micrOMEGAs} models are in general limited to real Lagrangian parameters, that is all couplings need to be real. This caused problems in our relic density calculations due to the presence of phases in multiple sectors. To overcome this, we maintained a full calculation including phases within {\tt SPheno} but recast the model with real valued couplings (taking the modulus) for the relic density calculation. As, in general, the $CP$-violating contribution to relevant (co-)annihilation channels are limited, and our $CP$-violating parameters are also numerically rather small, this is a valid approximation. We have verified for a few cases, that the effect of the phases had little impact on the amplitudes squared of the relevant processes.

Linking these tools together, we are able to investigate a wide range of constraints. The list of the Standard Model parameters to be fitted is given in Table \ref{Tab:SMConstraints} while the flavour and dark matter constraints are listed in Table \ref{Tab:LeptonConstraints}. We give the list of input parameters and their respective scanning range in Table \ref{Tab:input par}.

\begin{table}
	\centering
	\renewcommand{\arraystretch}{1.3}
	\scalebox{0.9}{
	\begin{tabular}{|c|c||c|c||c|c|}
		\hline
		Parameter & Range & Parameter & Range & Parameter & Range \\
		\hline
	 $A_0$ & $ [-1000, 1000]       $  &								$y_d$ & $ [ -2, 2 ]$ &			 		$k_2$ & $ [ -8, 8 ]$ \\	
	 $m_0^2$ & $ [2\cdot 10^5, 7\cdot 10^6]$ &						$z_1^d$ & $ [ -2, 2 ]$ &					$k_3$ & $ [ -8, 8 ]$ \\
	 $m_{H_u}^2$ & $ [2\cdot 10^5, 7\cdot 10^6]$ &					$x_2$ & $ [ -1.2, 1.2 ]$ &				$k_4$ & $ [ -8, 8 ]$ \\				
	 $m_{H_d}^2$ & $ [2\cdot 10^5, 7\cdot 10^6]$ &					$y_s$ & $ [ -1, 1 ]$ &                $B_0^{(N)}$ & $ [ 0, 8]$ \\	
	 $M_1$ & $ [500, 1500]$ &										$z_3^d$ & $ [ -2, 2 ]$ &		$B_3^{(N)}$ & $ [ -8, 8 ]$ \\
	 $M_2$ & $ [500, 1500]$ &		         		$z_2^d$ & $ [ -2, 2 ]$ &			$K_3^{(N)}$ & $ [ -8, 8 ]$ \\
	 $M_3$ & $ [800, 3000]$ &						$y_b$ & $ [ -2.5, 2.5 ]$ &				$a_u$ & $ [ -8, 8 ]$ \\	
	 $\tan(\beta)$ & $ [ 6, 15 ]$ &					$A$ & $ [ -8, 8]$ &		$a_c$ & $ [ -8, 8 ]$ \\
	 $y_u$ & $ [ 0.1, 1.5 ]$ &			    		$B$ & $ [ -8, 8 ]$ &				$a_t$ & $ [ -8, 8 ]$ \\	
	 $y_c$ & $ [ 0.1, 1.5 ]$ &						$C$ & $ [ -8. 8 ]$ &			$a_s$ & $ [ -8, 8 ]$ \\
	 $y_t$ & $ [ 0.4, 0.7 ]$ &						$D$ & $ [ -8, 8 ]$ &				$a_b$ & $ [ -8, 8 ]$ \\	
	 $\theta_3^d$ & $ [ 0, 2\pi]$ &					$b_{01}$ & $ [ 0, 8 ]$ &				$\tilde{x}_2^a$ & $ [ -8, 8 ]$ \\
	 $\theta_2^d$ & $ [ 0, 2\pi]$ &					$b_{02}$ & $ [ 0, 8 ]$  &			$z_3^{da}$ & $ [ -8, 8]$ \\	
	 $z_1^{da}$ & $ [ -8, 8 ]$ &					$b_2$ & $ [ -8, 8 ]$ &				$z_2^{da}$ & $ [ -8, 8]$ \\
	 $\alpha_D$ & $ [ -8, 8]$ &						$b_3$ & $ [ -8, 8 ]$ &			& \\
	 $y_D$ & $ [ -1.5, 1.5 ]$ &					 	$b_4$ & $ [ -8, 8 ]$ &		 &  \\

	\hline
	\end{tabular}
	}
	\caption{GUT scale input parameters for the model and their scanning range. For all parameters, the step size for a Markov chain iteration is given as 0.5\% of the total range length of the allowed values. This step size was procured by trial and error in order to balance scan efficiency and a search of the parameter space. In addition, we set a fixed value for the following parameters: ${\rm sign}(\mu) = -1$; and $\lambda = 0.22$ and $M_{\rm GUT} = 2\cdot 10^{16}$ which enters as parametrization of Yukawa, trilinear and mass matrices as stated in Eqs.\ \eqref{eqn:model_Yukawa_couplings}, \eqref{eqn:neutrino_couplings_UV} and \eqref{Eq:Soft_matrices}. }
	\label{Tab:input par}
\end{table}

\begin{table}
	\centering
	\renewcommand{\arraystretch}{1.3}
	\scalebox{0.9}{
	\begin{tabular}{|c|c|c|c|}
		\hline
		Observable & Constraint  & Refs. \\
		\hline
		$m_u$ & $(2.2\pm0.5)\cdot 10^{-3}$  & \cite{Zyla:2020zbs} \\
		$m_c$ & $1.275 \pm 0.0035$ & \cite{Zyla:2020zbs} \\
		$m_t$ & $172.76 \pm 0.9$ & \cite{Zyla:2020zbs}  \\
		$m_d$ & $(4.7\pm 0.5)\cdot 10^{-3}$ &  \cite{Zyla:2020zbs}  \\
		$m_s$ & $(93\pm 9)\cdot 10^{-3}$ &  \cite{Zyla:2020zbs}  \\
		$m_b$ & $4.18\pm 0.04$ &   \cite{Zyla:2020zbs}  \\
		$m_e$ & $0.511\cdot 10^{-3}$ & \cite{Zyla:2020zbs}  \\
		$m_\mu$ & $105.66\cdot 10^{-3}$ & \cite{Zyla:2020zbs}  \\
		$m_\tau$ & $1.7769$ &  \cite{Zyla:2020zbs}  \\
		$m_h$ & $125$  &  \cite{Zyla:2020zbs}  \\
		\hline
	\end{tabular}
	}	\renewcommand{\arraystretch}{1.3}
	\scalebox{0.9}{
	\begin{tabular}{|c|c|c|c|}
		\hline
		Observable & Constraint  & Refs. \\
		\hline
		 $(\Delta m_{21}^{\nu})^2$ & $(7.42 \pm 0.2)\cdot 10^{-23}$ &    \cite{Esteban:2018azc}  \\
	    $(\Delta m_{31}^{\nu})^2$ & $(2.514\pm 0.028)\cdot 10^{-21}$ &    \cite{Esteban:2018azc}  \\
		$\sin(\theta_{12}^{\text{CKM}})$ & $0.225 \pm 0.0010$ &    \cite{UTfit}  \\
		$\sin(\theta_{13}^{\text{CKM}})$ & $(0.003675 \pm 9.5) \cdot 10^{-5}$ &    \cite{UTfit}  \\
		$\sin(\theta_{23}^{\text{CKM}})$ & $0.042 \pm 0.00059$ &    \cite{UTfit}  \\
		$\delta^{\text{CKM}}$ & $1.168 \pm 0.04$ &    \cite{UTfit}  \\
		$\sin(\theta_{12}^{\text{PMNS}})$ & $0.55136 \pm 0.012$ &    \cite{Esteban:2018azc}  \\
		$\sin(\theta_{13}^{\text{PMNS}})$ & $0.1490 \pm 0.0022$    & \cite{Esteban:2018azc}  \\
		$\sin(\theta_{23}^{\text{PMNS}})$ & $0.7550 \pm 0.0134$    & \cite{Esteban:2018azc}  \\
		$\delta^{\text{PMNS}}$ & $3.86 \pm 1.2$ &    \cite{Esteban:2018azc}  \\
		\hline
	\end{tabular}
	}
	\caption{$\nu$SM parameters, masses and EWSB constraints for our model exploration. All masses are given in GeV and are pole masses, except for the bottom and light quarks: the bottom (light quarks) one is the $\overline{\text{MS}}$ mass given at the scale $Q = m_b$ ($\mu = 2$ GeV). Theoretical uncertainties of $1\%$ are assumed for the different masses and are added in quadrature with the experimental ones. Note that the charged lepton and Higgs boson mass experimental uncertainties are negligible with respect to the theoretical ones and are therefore omitted.}
	\label{Tab:SMConstraints}
\end{table}

\begin{savenotes}
\begin{table}
	\centering
	\renewcommand{\arraystretch}{1.2}
	\scalebox{0.85}{
	\begin{tabular}{|c|c|c|}
		\hline
		Observable & Constraint  & Refs.  \\  
		\hline

		$\mathrm{BR}(\mu \rightarrow e\gamma)$ & $ <4.2 \cdot 10^{-13}$ &     \cite{Zyla:2020zbs}  \\
		$\mathrm{BR}(\tau \rightarrow e\gamma)$ & $ <3.3 \cdot 10^{-8}$ &     \cite{Zyla:2020zbs}  \\
		$\mathrm{BR}(\tau \rightarrow \mu\gamma)$ & $ <4.4 \cdot 10^{-8}$ &     \cite{Zyla:2020zbs} \\
		$\mathrm{CR}(\mu - e,Ti)$ & $<4.3\cdot 10^{-12}$ &     \cite{Zyla:2020zbs}  \\
		$\mathrm{CR}(\mu - e,Au)$ & $<7\cdot 10^{-13}$ &     \cite{Zyla:2020zbs}  \\
		$\mathrm{CR}(\mu - e,Pb)$ & $<4.6\cdot 10^{-11}$ &     \cite{Zyla:2020zbs}  \\
		$\mathrm{BR}(\mu \rightarrow 3e)$ & $ <1\cdot 10^{-12}$     & \cite{Zyla:2020zbs}  \\
		$\mathrm{BR}(\tau \rightarrow 3e)$ & $ <2.7\cdot 10^{-8}$     & \cite{Zyla:2020zbs}  \\
		$\mathrm{BR}(\tau \rightarrow 3\mu)$ & $ <2.1\cdot 10^{-8}$     & \cite{Zyla:2020zbs}  \\
		$\mathrm{BR}(\tau- \rightarrow e^-\mu^+\mu^-)$ & $ <2.7\cdot 10^{-8}$     & \cite{Zyla:2020zbs}  \\
		$\mathrm{BR}(\tau- \rightarrow \mu^-e^+\mu^-)$ & $ <1.8\cdot 10^{-8}$     & \cite{Zyla:2020zbs}  \\
		$\mathrm{BR}(\tau- \rightarrow e^+\mu^-\mu^-)$ & $ <1.7\cdot 10^{-8}$     & \cite{Zyla:2020zbs}  \\
		$\mathrm{BR}(\tau- \rightarrow \mu^+e^-e^-)$ & $ <1.5\cdot 10^{-8}$     & \cite{Zyla:2020zbs} \\
        \hline
	\end{tabular}
	}	\renewcommand{\arraystretch}{1.2}
	\scalebox{0.85}{
	\begin{tabular}{|c|c|c|}
		\hline
		Observable & Constraint  & Refs. \\ 
		\hline
		  $\mathrm{BR}(Z \rightarrow e\mu)$ & $ <7.5\times10^{-7}$ & \cite{Zyla:2020zbs}  \\
		$\mathrm{BR}(Z \rightarrow e\tau)$ & $ <9.8\times10^{-6}$ &     \cite{Zyla:2020zbs}  \\
		$\mathrm{BR}(Z \rightarrow \mu\tau)$ & $ <1.2\times10^{-5}$ &     \cite{Zyla:2020zbs}  \\
		$\mathrm{BR}(h \rightarrow e\mu)$ & $ <6.1\times10^{-5}$ &     \cite{Zyla:2020zbs}  \\
		$\mathrm{BR}(h \rightarrow e\tau)$ & $ <4.7\times10^{-3}$ &     \cite{Zyla:2020zbs}  \\
		$\mathrm{BR}(h \rightarrow \mu\tau)$ & $ <2.5\times10^{-3}$ &     \cite{Zyla:2020zbs}  \\
		$\mathrm{BR}(\tau \rightarrow e\pi)$ & $ <8\times10^{-8}$ &     \cite{Zyla:2020zbs}  \\
		$\mathrm{BR}(\tau \rightarrow e\eta)$ & $ <9.2\times10^{-8}$ &     \cite{Zyla:2020zbs}  \\
		$\mathrm{BR}(\tau \rightarrow e\eta')$ & $ <1.6\times10^{-7}$ &     \cite{Zyla:2020zbs}  \\
		$\mathrm{BR}(\tau \rightarrow \mu\pi)$ & $ <1.1\times10^{-7}$ &     \cite{Zyla:2020zbs}  \\
		$\mathrm{BR}(\tau \rightarrow \mu\eta)$ & $ <6.5\times10^{-8}$ &     \cite{Zyla:2020zbs}  \\
		$\mathrm{BR}(\tau \rightarrow \mu\eta')$ & $ <1.3\times10^{-7}$ &     \cite{Zyla:2020zbs}  \\

		\hline
		\hline
		$\Omega h^2$  & \begin{tabular}{@{}c@{}}$ 0.12 \pm 0.012$ th. \end{tabular} & \cite{Ade:2015xua, Belanger:2001fz,  Belanger:2004yn, Barducci:2016pcb} \\
        \hline
        Direct detection & cf. Figure \ref{fig:DMDD} & \cite{Aprile:2018dbl, Aprile:2019dbj}\\
		\hline
	\end{tabular}
	}
	\caption{Leptonic flavour and dark matter constraints. These upper limits numbers are given at the 90\% confidence level. For the dark matter relic density we assume 10\% theoretical uncertainties because of cosmological assumptions. } 
	\label{Tab:LeptonConstraints}
\end{table}
\end{savenotes}

In all tables, the upper bounds constraints are given at the 90\% confidence level and in order to help the process of chain convergence and initialization, we postulate a smoothing step function for the upper limits constraints 
\begin{equation}
    \mathcal{L}_{\text{upper}}(\vec{\theta},O_i, \sigma_i) = \left\{
                \begin{array}{ll}
                  1  \hspace*{3cm} \text{for}\ \ O_i(\vec{\theta}) \leq O_{i}^{\text{bound}} \\
                  \ \\
                  e^{-\frac{\left(O_{i}(\Vec{\theta})-O_{i}^{\text{bound}}\right)^2}{2 \sigma_i^2}}\ \ \ \  \text{for} \ \ \  O_i(\Vec{\theta}) >  O_{i}^{\text{bound}}
                \end{array}
              \right.
\end{equation}
where we chose, somehow arbitrarily, a common value of $\sigma_i = 10\% \cdot O_{i}^{\text{bound}}$. 

On the other hand, we associate a Gaussian likelihood function for all experimentally measured observables
\begin{equation}
 \mathcal{L}_{\text{measured}}(\vec{\theta},O_i, \sigma_i)  =  e^{-\frac{\left(O_{i}(\Vec{\theta})-O_{i}^{\text{exp}}\right)^2}{2\sigma_{i}^2}}\,,
\end{equation}
$\sigma_i$ being the uncertainty given in Tables \ref{Tab:SMConstraints} and \ref{Tab:LeptonConstraints}.

\begin{figure}
    \centering
    \includegraphics[scale = 0.6]{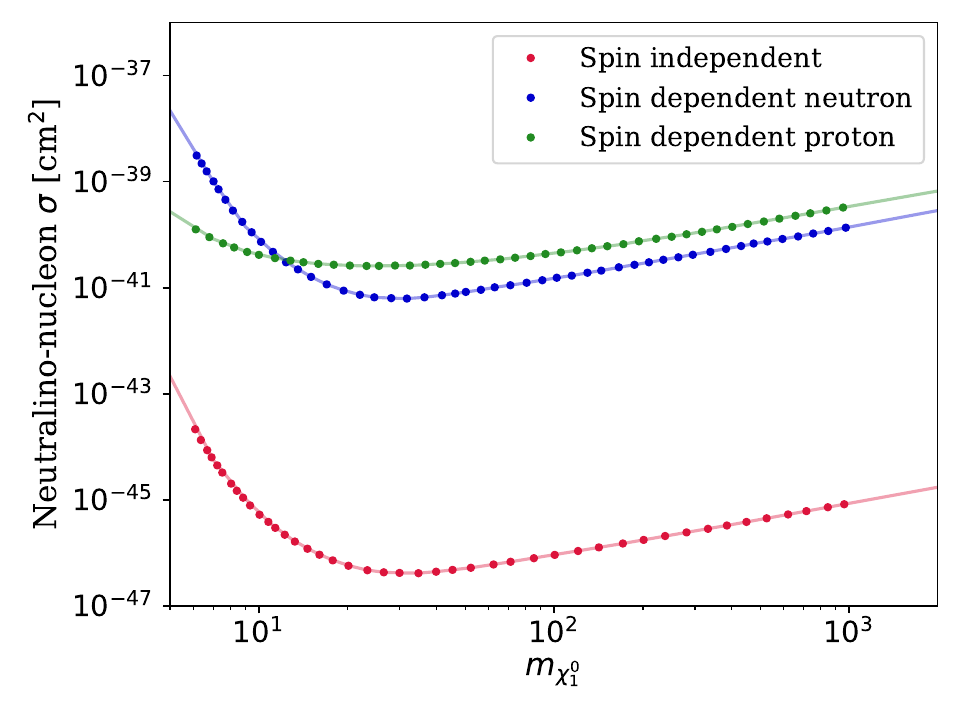}
    \caption{Dark matter direct detection limits in plane of dark matter mass and spin-(in)dependent nucleon scattering cross section. The dots correspond to data extracted from Refs.\ \cite{Aprile:2018dbl, Aprile:2019dbj}, while the solid line is the extrapolation we performed.}
    \label{fig:DMDD}
\end{figure}

Regarding the dark matter direct detection constraints, we extracted and extrapolated the curves from Refs.\ \cite{Aprile:2018dbl, Aprile:2019dbj} as shown in Figure \ref{fig:DMDD}, while for the relic density we have used the results from Ref.\ \cite{Ade:2015xua} adding a 10\% uncertainty due to {\tt micrOMEGAs} precision in combination with underlying cosmological assumptions. 
For the other constraints we are using the current experimental uncertainties associated to the values given in the different tables while adding in quadrature a theoretical constraints on the different standard model masses. The theory uncertainty on the Higgs mass is fixed at 2 GeV \cite{Slavich:2020zjv}
while we assume a common 1\% uncertainty on the different fermion masses because of RGE fixed order precision and changes from the $\overline{\text{DR}}$ to the on-shell renormalisation scheme. If no experimental constraints is present for a given value in the tables it is understood that theoretical uncertainties are by far dominant with respect to the experimental ones. Finally, the different quark masses are extracted at different scales, i.e.\ $Q = 2$ GeV for $q = (u,\, d,\, c,\, s)$ and $Q = m_b$ for $q=b$.

\begin{figure}
\centering
\makeatletter
\@for\sun:={omegaDM_hist,mh}\do{
\includegraphics[width=.45\textwidth]{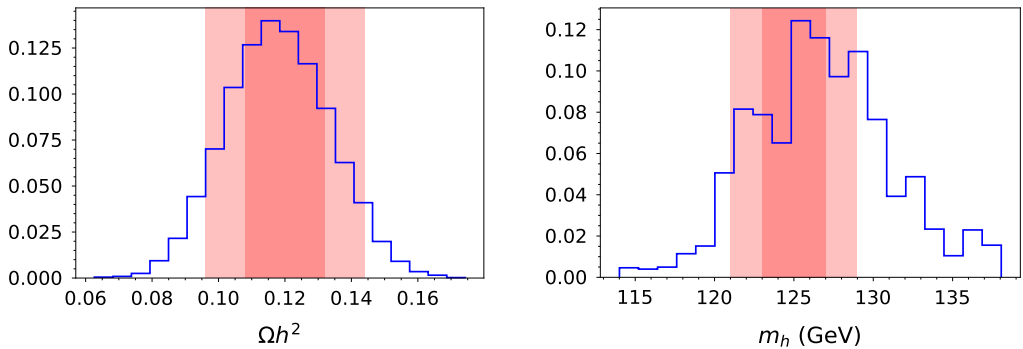}
\makeatother}
\caption{Distribution of the Higgs mass and dark matter relic density predictions, normalised to one. The $1\sigma$ ($2\sigma$) region is marked in red (light red). The MCMC displays an approximate Gaussian fit around the experimental values.}
\label{fig:example constraints}
\end{figure}

Having implemented and executed the above, 197 chains were recovered. As the parameter space was so vast and such a large number of very precise constraints were used, the efficiency of these scans was very low, requiring weeks of computer time to complete. Therefore, the scans were allowed 2000 Markov chain steps each. After this process we collect the data and applied a likelihood cutoff such that only points with relatively high likelihood are left in the final data set. This was in order to prevent the distributions presented here-in from misleading the reader into thinking some parts of parameter space were viable when, in fact, they produce excessively low likelihoods. The likelihood cutoff applied was $10^{-150}$. Although this is tiny, much of the poor likelihood comes from poor convergence of the fermion masses (see discussion in Section \ref{sec:results fermions}). In general, the remaining constraints converged very well to the observed values. As an example of two constraints, see Figure \ref{fig:example constraints}, which shows how the Higgs boson mass and relic density are centered around the expected value.
\section{Results}
\label{sec:results}

The subsequent results are based on the Markov Chain Monte Carlo (MCMC) study following the methods elucidated previously. Having already presented two illustrative plots showing the constraints used to guide the MCMC, we now present the resultant spectra and phenomena. We begin with a discussion of the fermion masses, mixing, and a general discussion of the model's success in recreating the Standard Model observables. We then look at the supersymmetric (SUSY) spectrum, the dark matter sector, and further phenomenological results. Finally, we give a discussion of the effects on collider physics and experimental physics more generally. 

\subsection{Fermion masses and mixing \label{sec:results fermions}}

\begin{figure}[!ht]
    \centering
    \includegraphics[scale = 0.29]{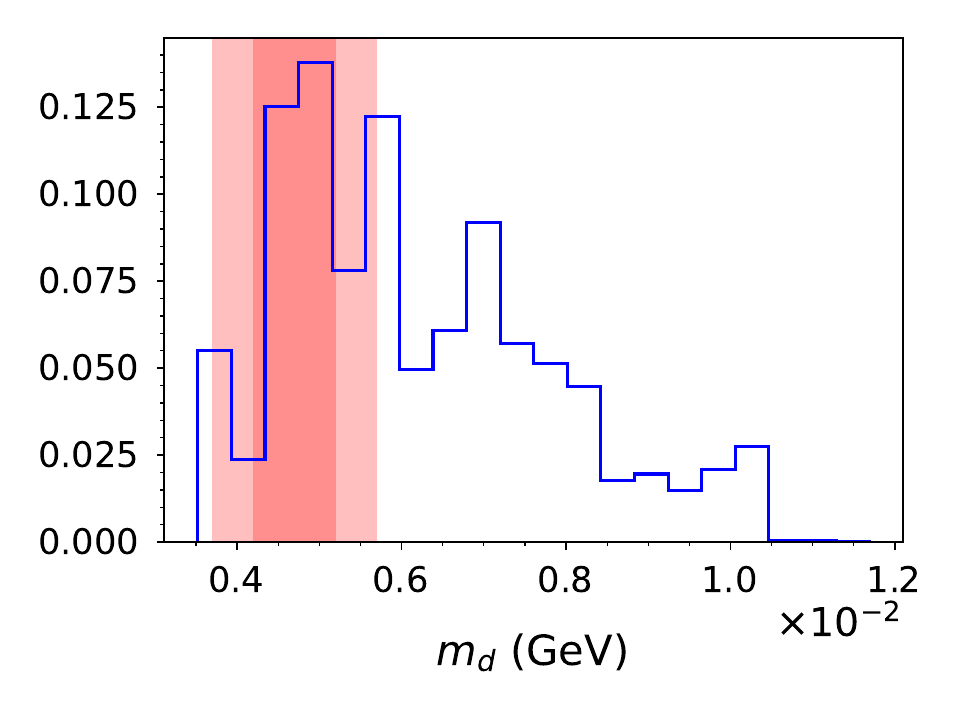} \ \ \includegraphics[scale = 0.29]{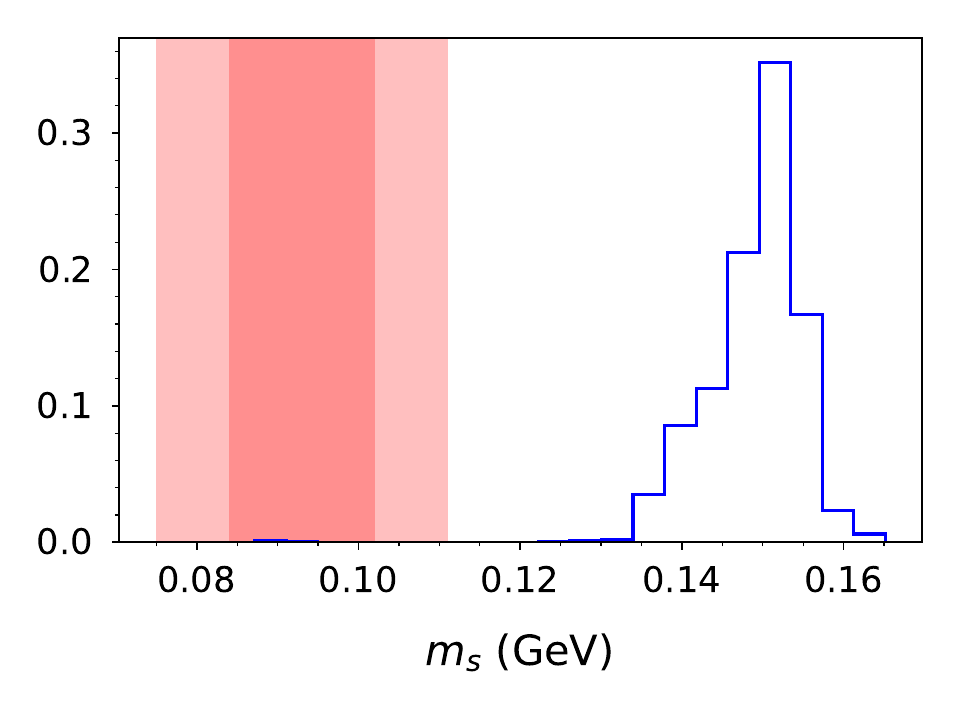} \ \ \includegraphics[scale = 0.29]{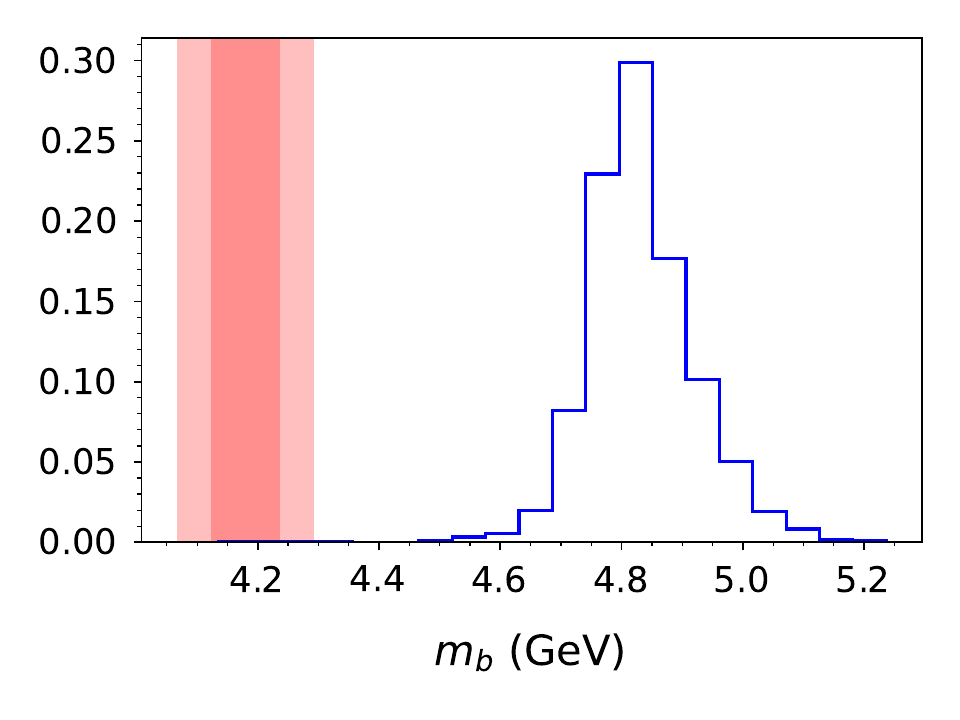}
    \includegraphics[scale = 0.29]{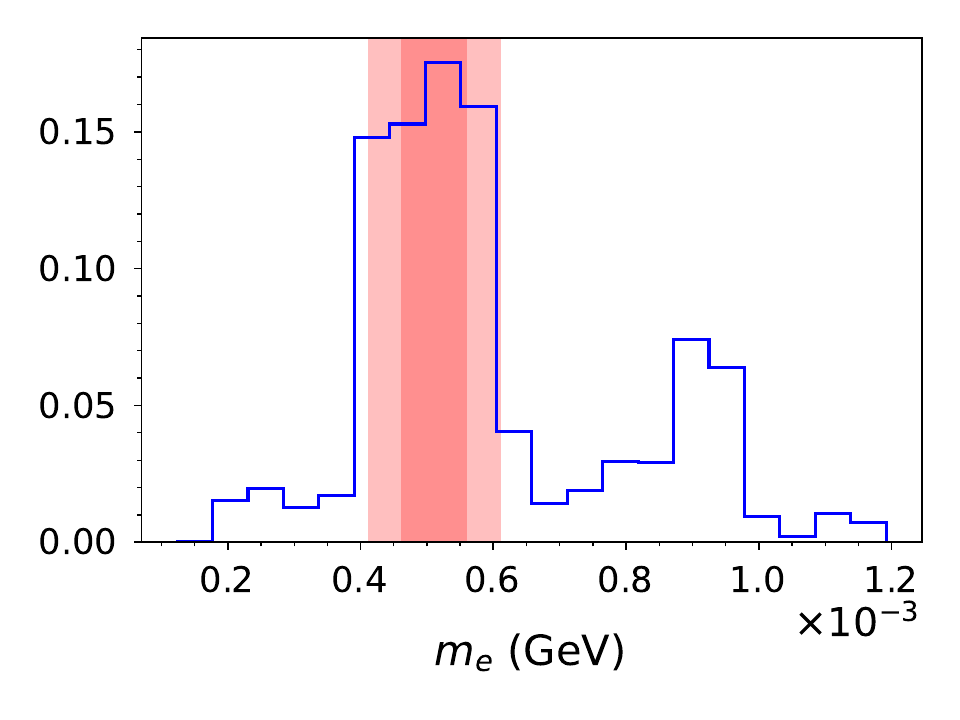} \ \ \includegraphics[scale = 0.29]{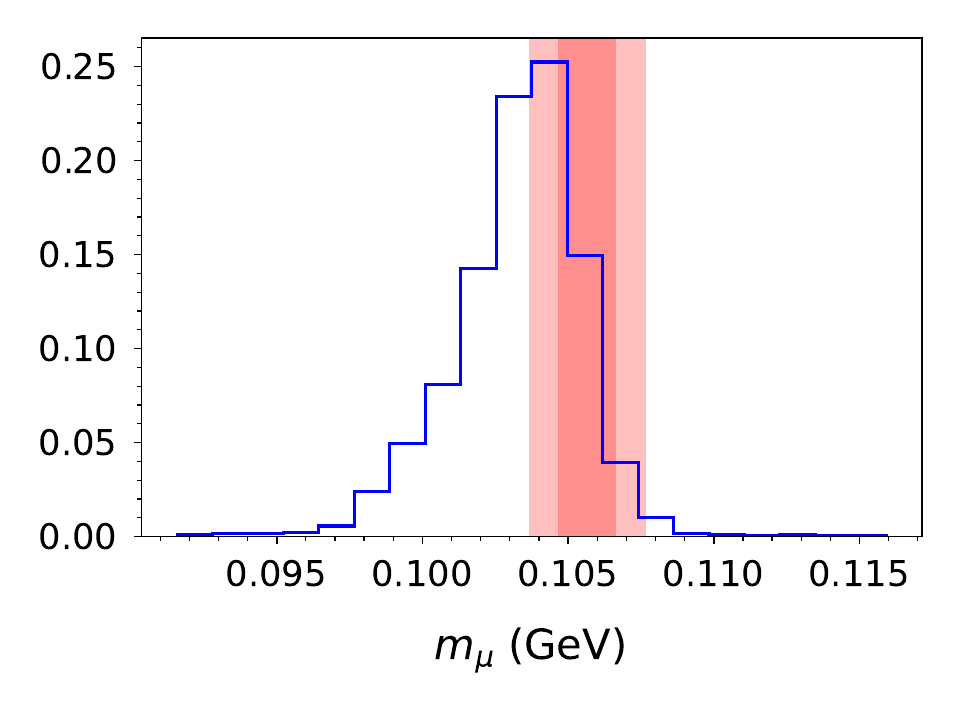} \ \ \includegraphics[scale = 0.29]{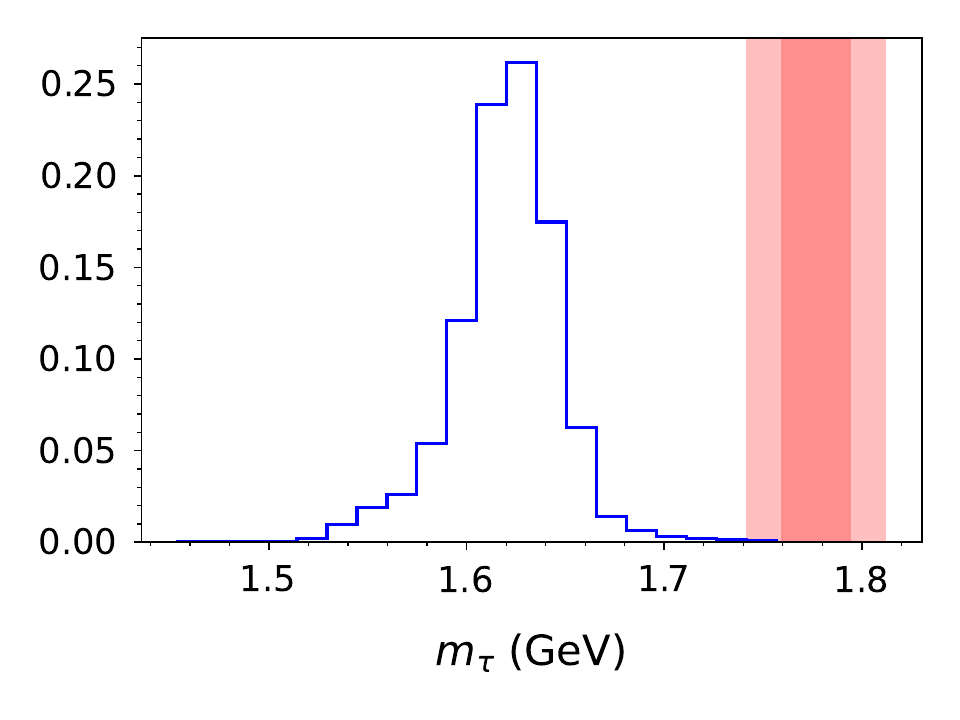}
    \includegraphics[scale = 0.29]{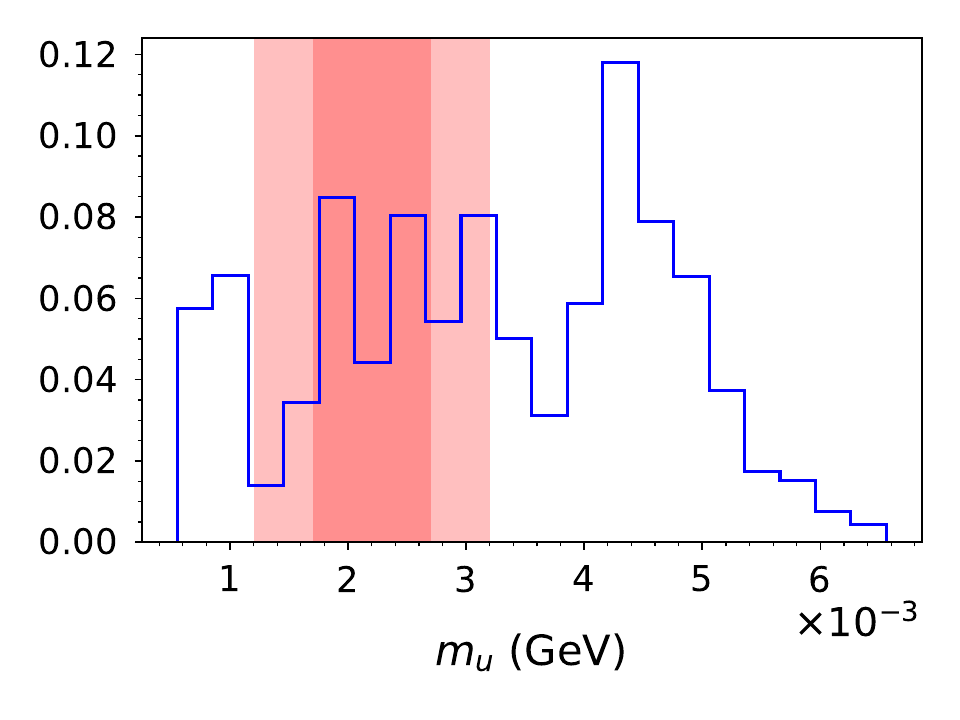} \ \ \includegraphics[scale = 0.29]{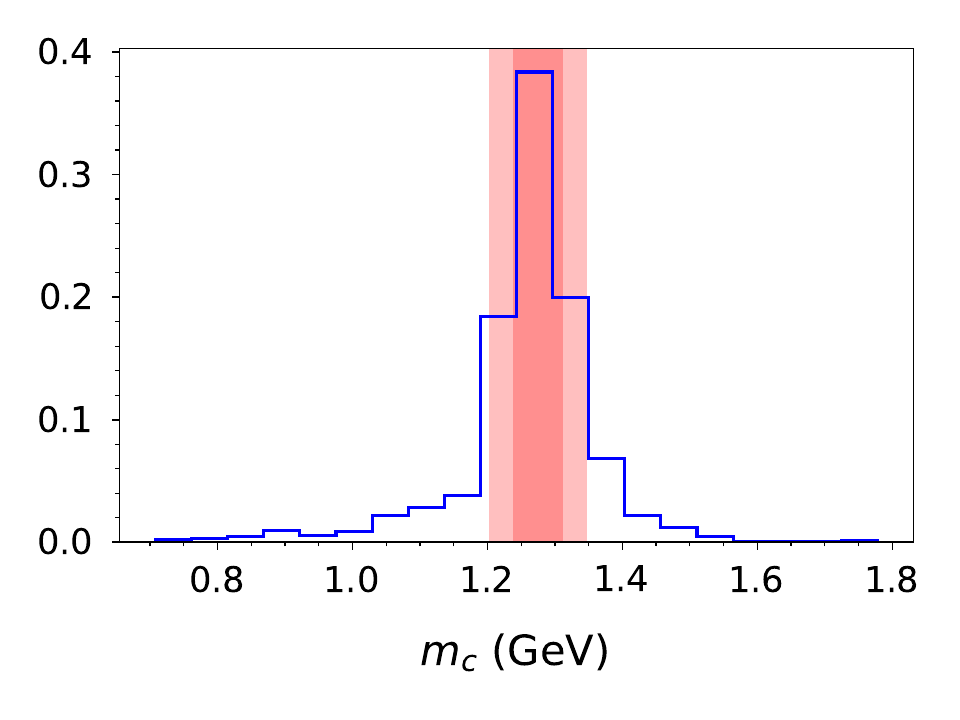} \ \ \includegraphics[scale = 0.29]{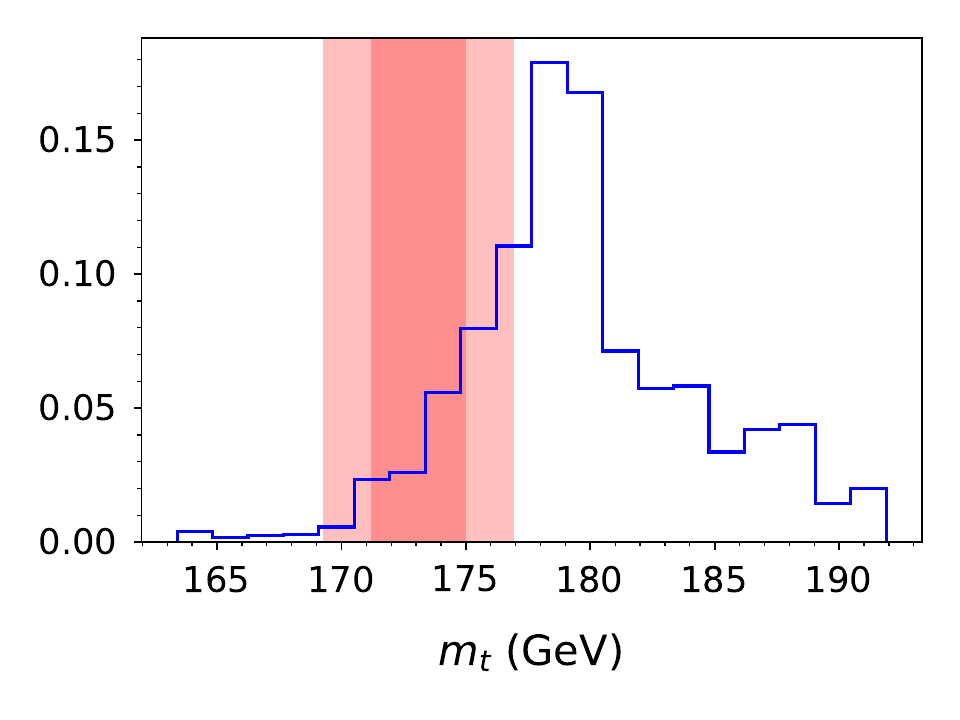}
    \includegraphics[scale = 0.29]{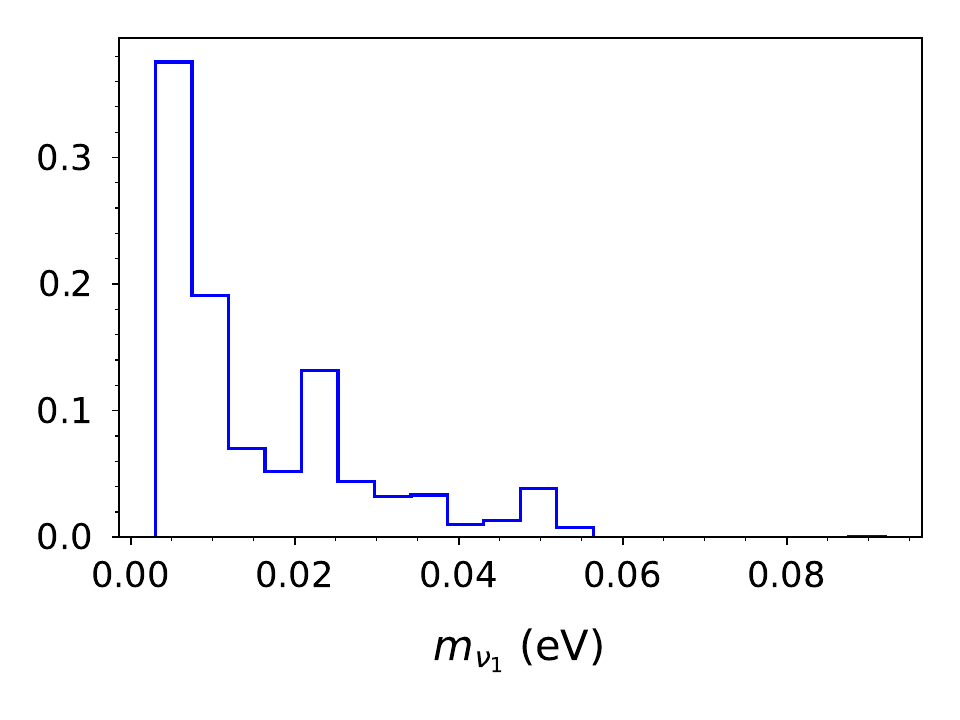} \ \ \includegraphics[scale = 0.29]{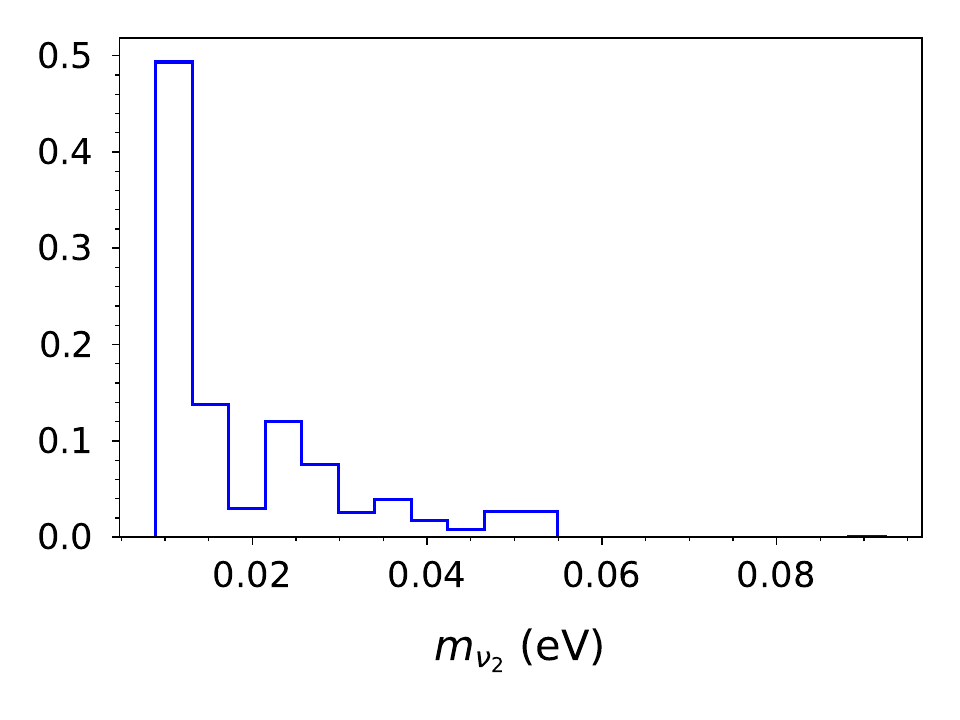} \ \ \includegraphics[scale = 0.29]{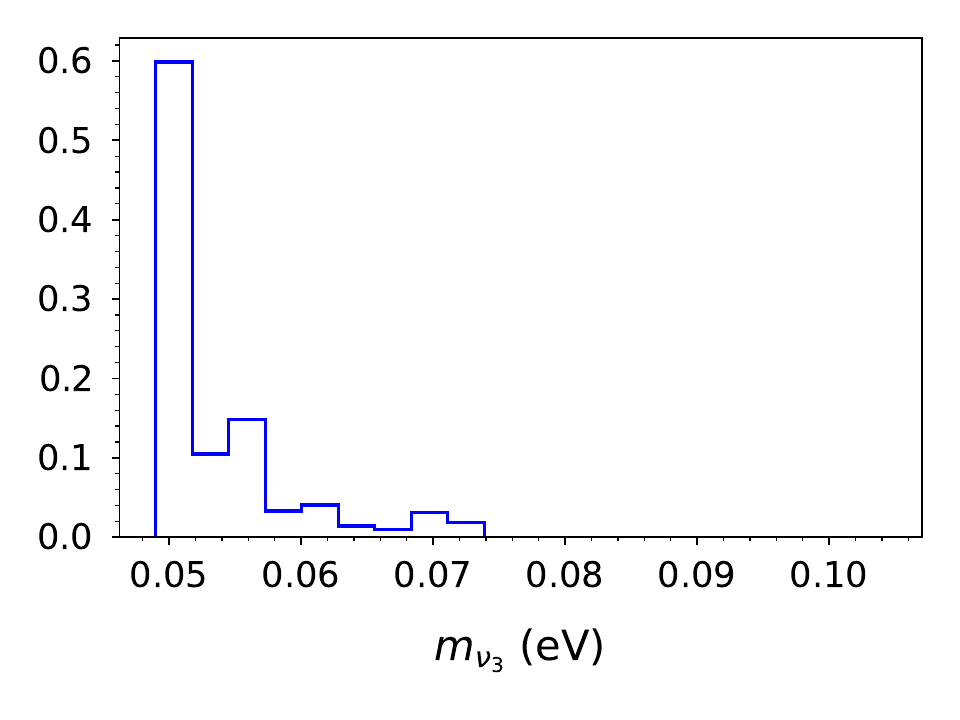}
    \caption{The fermion masses are displayed where the red (light red) region indicating the $1\sigma$ ($2\sigma$) limits. The first generations are very well fitted. However, due to the link between the down type quarks and the leptons at the GUT scale, the second and third generation masses are slightly off. Furthermore, the top mass is also slightly poorly aligned due to the Higgs mass constraint. Note that few points exhibit neutrino masses above the visible part of the histograms.}
    \label{fig:masses}
\end{figure}

\begin{figure}
    \centering
    \includegraphics[scale = 0.43]{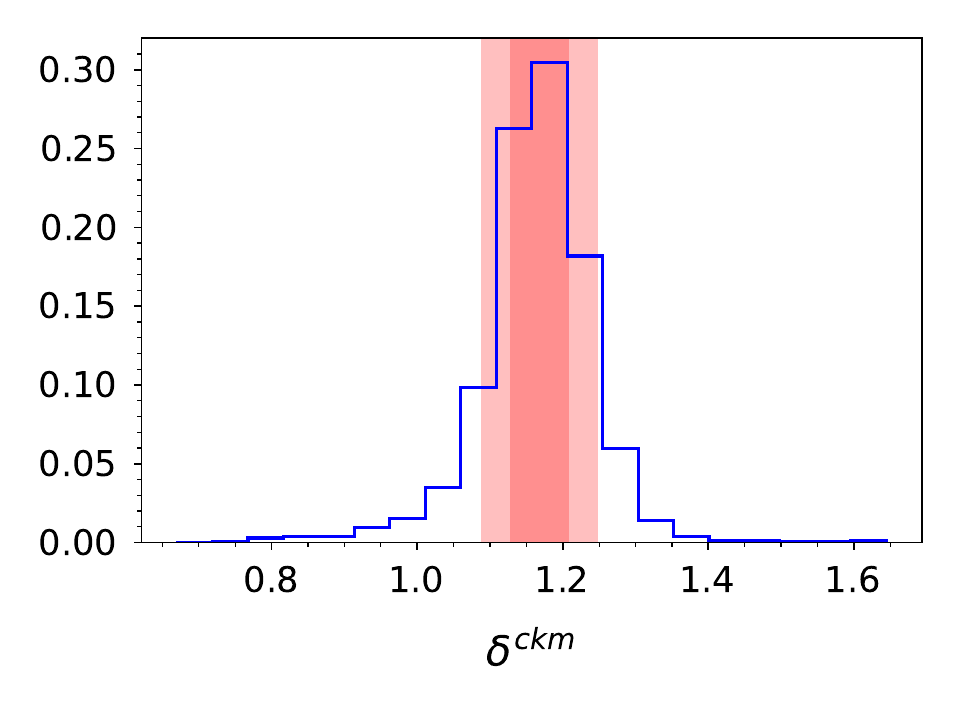} 
    \includegraphics[scale = 0.43]{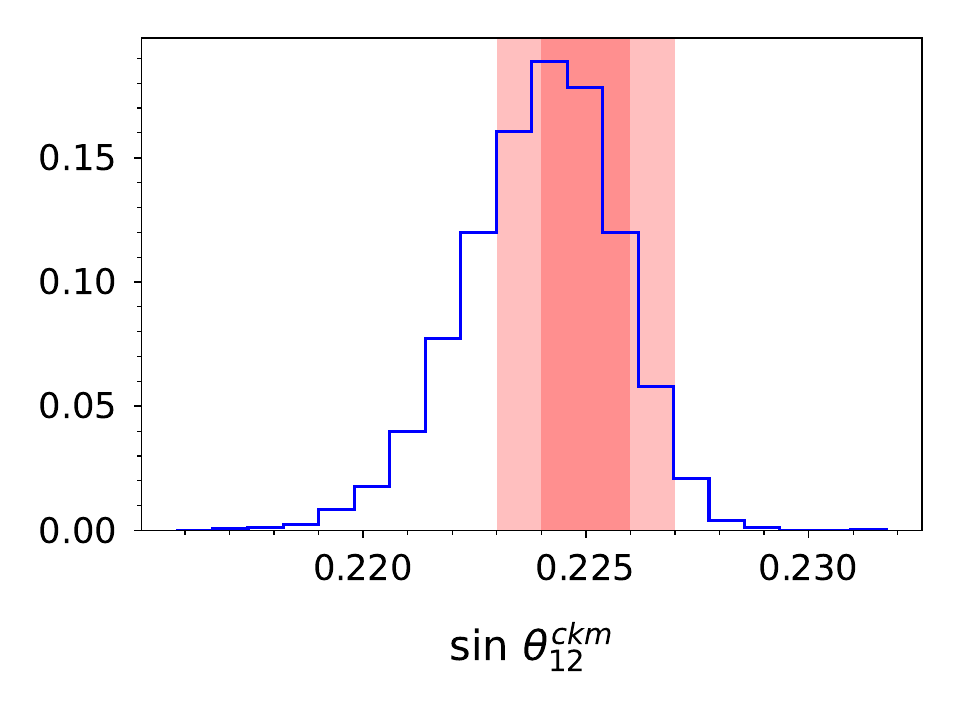} \\
    \includegraphics[scale = 0.43]{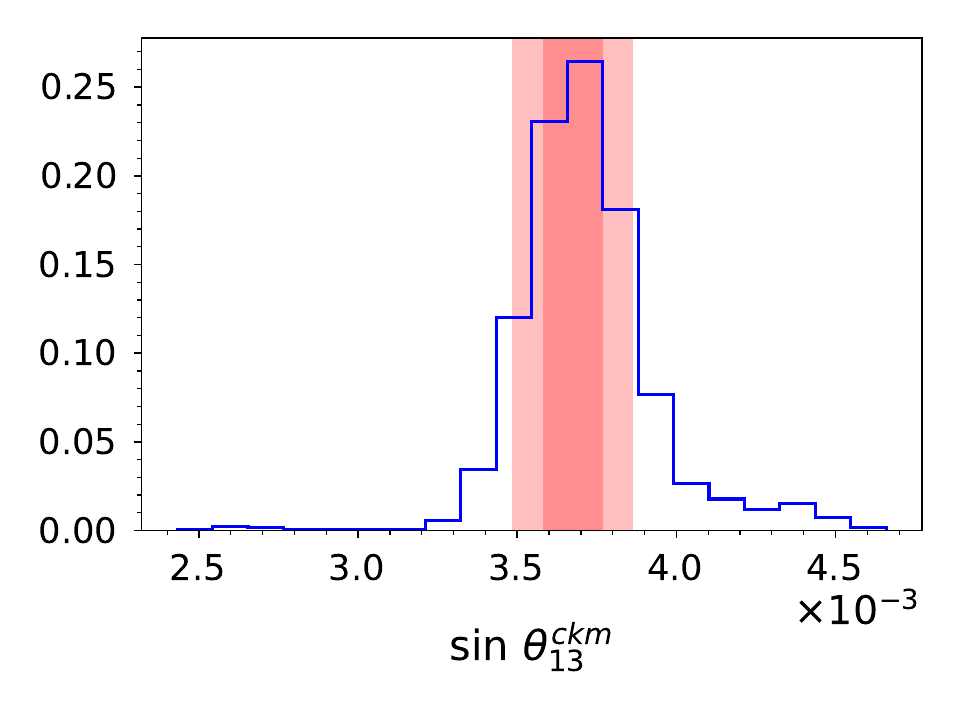}  
    \includegraphics[scale = 0.43]{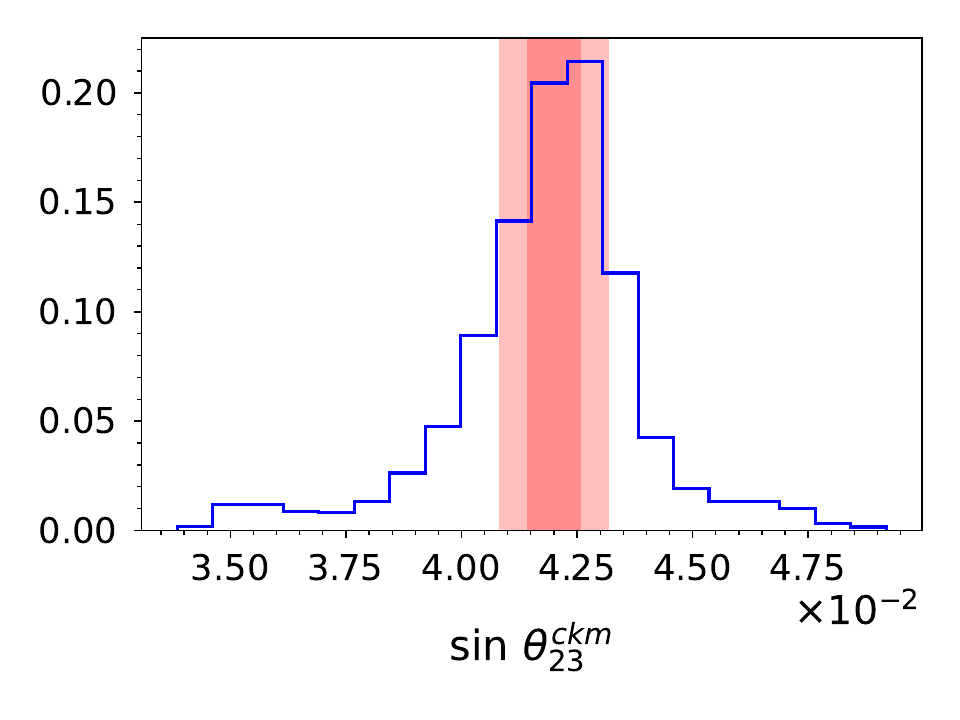}
    \caption{The CKM parameters are displayed where the red (light red) region indicating the $1\sigma$ ($2\sigma$) limits. All parameters in the CKM are fitted very well with an approximately Gaussian distribution around the expected value.}
    \label{fig:mixingsCKM}
\end{figure}

\begin{figure}
    \centering
    \includegraphics[scale = 0.43]{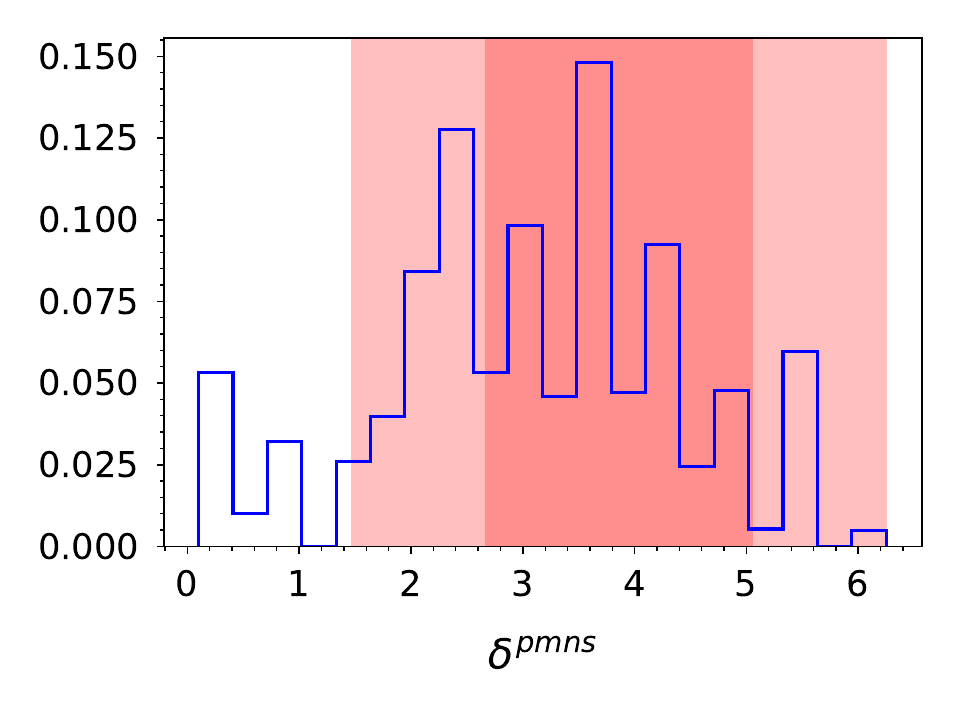} 
    \includegraphics[scale = 0.43]{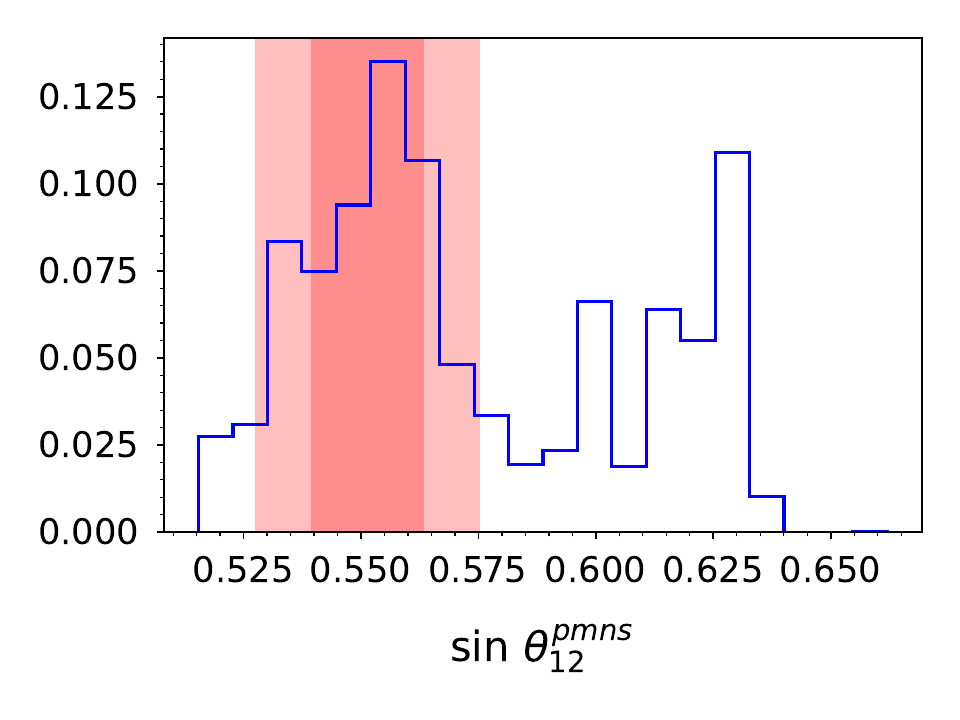} \\ 
    \includegraphics[scale = 0.43]{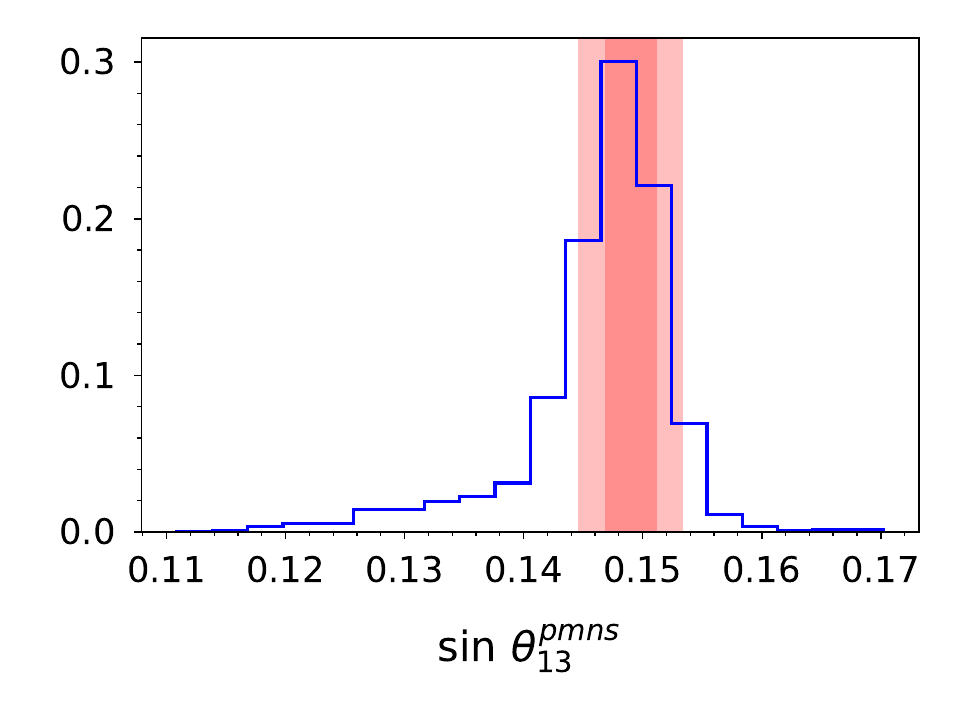} 
    \includegraphics[scale = 0.43]{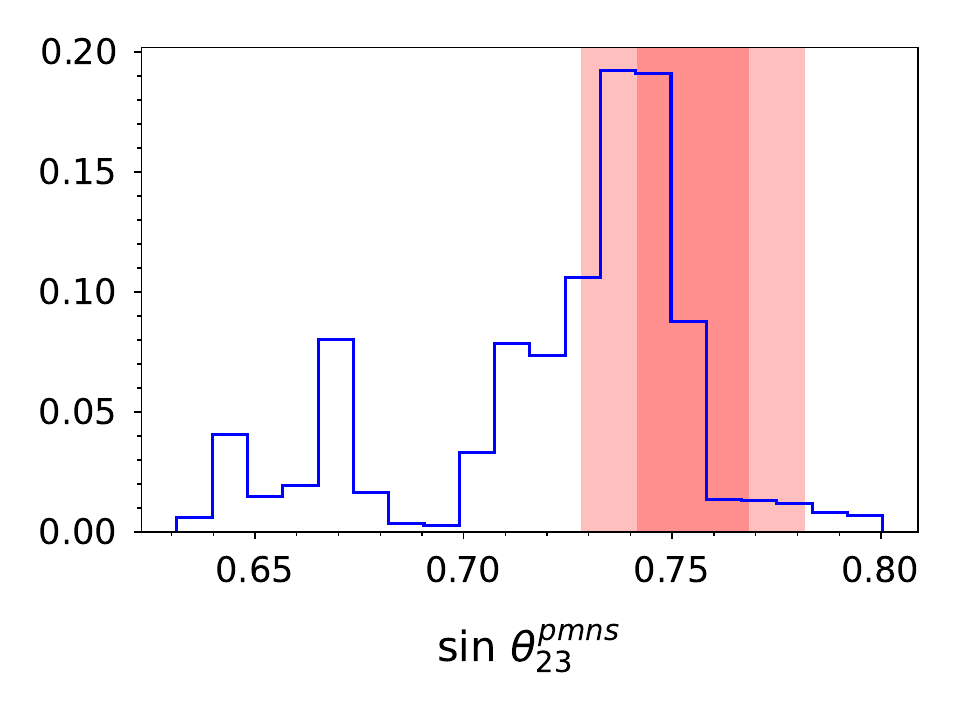}
    \caption{The PMNS parameters are displayed where the red (light red) region indicating the $1\sigma$ ($2\sigma$) limits. Unlike the CKM parameters, the PMNS shows more variation due to the less stringent experimental constraints.}
    \label{fig:mixingsPMNS}
\end{figure}

In this subsection we present the results of our scan for fermion masses and mixing parameters, which are put in as constraints as shown in Table \ref{Tab:SMConstraints}. The results for the fermion masses are shown in Figure \ref{fig:masses}, while those for the mixing parameters are shown in Figures \ref{fig:mixingsCKM} and \ref{fig:mixingsPMNS}. 

These results follow from the charged fermion Yukawa matrices at the GUT scale shown in Eqs.\ \eqref{eqn:model_Yukawa_couplings},
together with the neutrino Dirac Yukawa matrix in Eq.\ \eqref{Dirac} and the heavy right-handed neutrino mass matrix in 
Eq.\ \eqref{eqn:neutrino_couplings_UV}. Note that the $(3,3)$ entries of the charged lepton and down type quark Yukawa matrices 
are equal at the GUT scale (yielding approximate bottom-tau unification $m_b=m_s$), while the $(2,2)$ entries of these matrices 
differ by the Georgi-Jarlskog (GJ) factor of $3$ (yielding an approximate strange to muon mass ratio $m_s = m_{\mu}/3$ at the GUT scale).

The results for fermion masses in Figure \ref{fig:masses} show that the above GJ relations do not lead to phenomenologically viable charged lepton and down type quark masses at low energy, in particular $m_s$, $m_b$ and $m_{\tau}$ are not well fitted.
This problem has also been noted by other authors, and possible solutions have been proposed based on various alternative choices of GUT scale Higgs leading to different phenomenologically successful mass ratios at the GUT scale \cite{Antusch:2009gu, Antusch:2013rxa}. Since the purpose of this paper is to perform a comprehensive phenomenological analysis on an existing benchmark model, we shall not consider such alternative solutions here, but simply note that such solutions exist and could be readily applied.

\begin{figure}
    \centering
    \includegraphics[scale = 0.7]{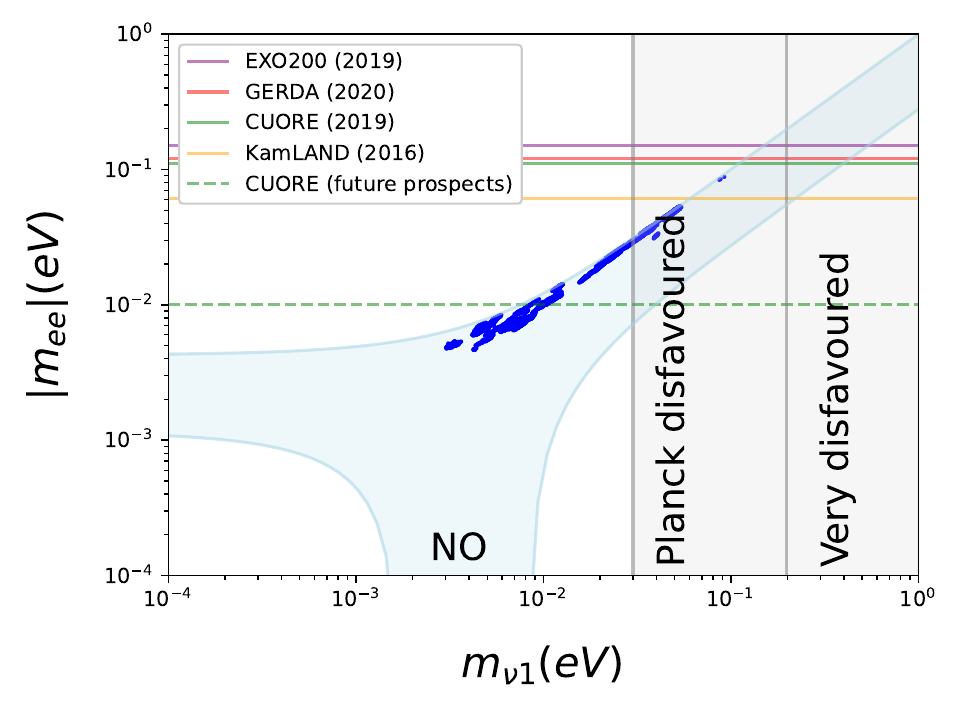} 
    \caption{The dark blue points are the model prediction of the neutrinoless double beta decay parameter $|m_{ee}|$ vs the mass of the lightest neutrino $m_{\nu 1}$. The light blue shaded region shows the allowed range in this plane for a normal hierarchy as predicted by the model. The vertical grey shaded bands to the right show the current Planck disfavoured region \cite{Planck:2018vyg}, while the coloured horizontal lines show the limits on $|m_{ee}|$ from KamLAND \cite{KamLAND-Zen:2016pfg}, EXO-200 \cite{EXO-200:2019rkq}, CUORE \cite{CUORE:2019yfd}, and GERDA \cite{GERDA:2020xhi}. We also indicate future prospects for CUORE \cite{CUORE:2017fp}.}
    \label{fig:0nuBB}
\end{figure}

Note that the absolute values of the neutrino masses in Figure \ref{fig:masses} are genuine predictions of the model, since only the experimentally measured mass squared differences in Table~\ref{Tab:SMConstraints} were put in as constraints. 
In particular, the lightest neutrino mass distribution is peaked around a few times $10^{-3}$ eV. This leads to an interesting prediction for neutrinoless double-beta becay. In Figure \ref{fig:0nuBB} we give the model prediction of the neutrinoless double-beta decay parameter $m_{ee}$ against the mass of the lightest neutrino $m_{\nu 1}$. $m_{ee}$ is given by
\begin{equation}
    |m_{ee}| = |\sum_{i=1}^3 {U_{1i}}^2 m_{\nu i}| \,,
\end{equation}
where $m_{\nu i}$ are the light neutrino masses and $U$ is the PMNS matrix including the Majorana phases. We use the same convention as in Ref.\ \cite{Zyla:2020zbs}. Future projections for CUORE rule out approximately $52\%$ of the data set. Indeed, the model tends to favour relatively high values of $m_{ee}$ as compared to the theoretically allowed region. 

\begin{figure}
    \centering
    \includegraphics[scale = 0.43]{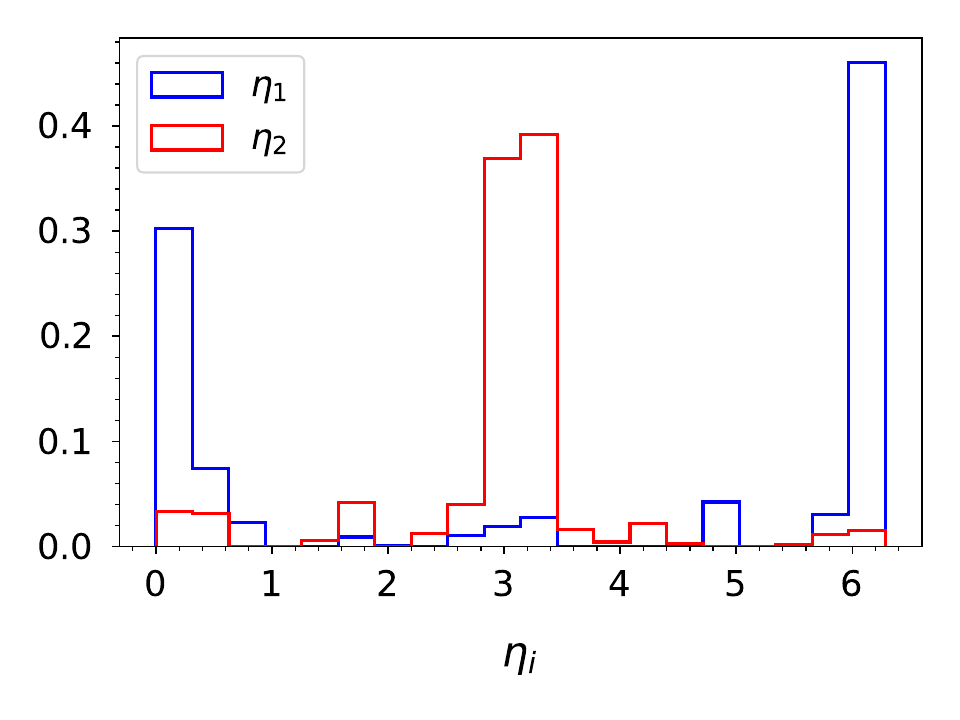}
    \includegraphics[scale = 0.43]{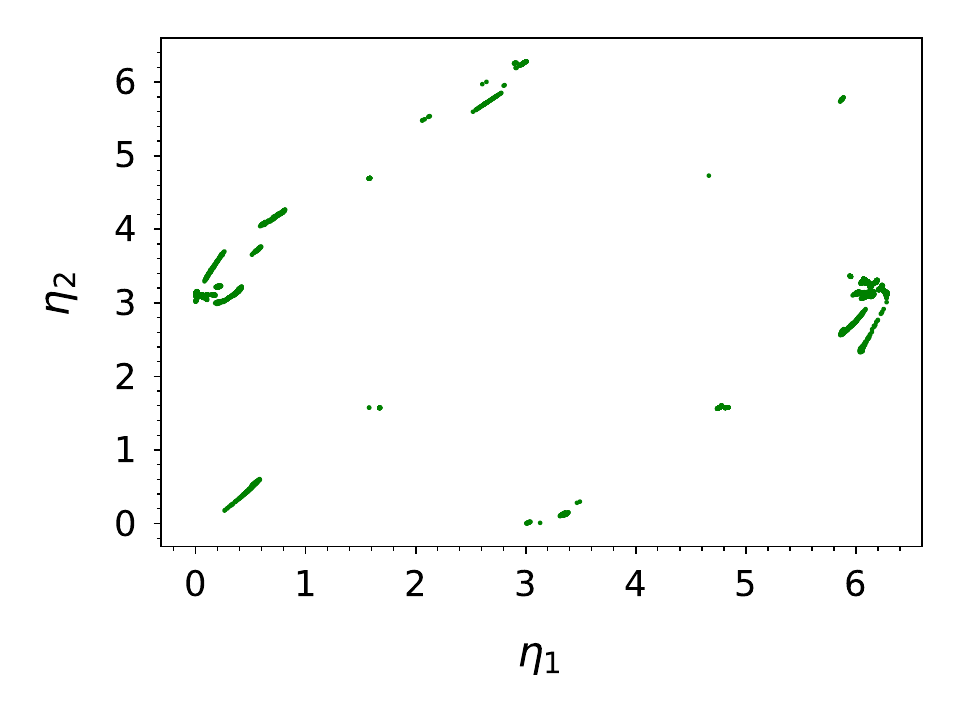}
    \caption{A representation of the Majorana phases is presented. The first panel shows the density of values for the two phases while the second shows their strong correlations.  }
    \label{fig:majorana phases}
\end{figure}

The results for the CKM and PMNS mixing parameters in Figures \ref{fig:mixingsCKM} and \ref{fig:mixingsPMNS} show a good fit to the constraints. The PMNS mixing parameters  $\sin \theta_{12}^{\rm PMNS}$ and $\sin \theta_{13}^{\rm PMNS}$ also fit very well. However the model prefers somewhat smaller values of $\sin \theta_{23}^{\rm PMNS}$, with the $CP$-oscillation phase $\delta^{\rm PMNS}$ being quite uniformly distributed.

The Majorana phases are also predicted and are highly correlated as shown in Figure \ref{fig:majorana phases}. In principle, the fact that they are correlated is not surprising. The model only depends on two high scale phases, $\theta_2^d$ and $\theta_3^d$, and the MCMC has two low scale constraints on the phases from $\delta^{CKM}$ and $\delta^{PMNS}$. Therefore, the high scale phases, who determine the Majorana phases, must be correlated. However, the striking nature of the correlation is surprising. It seems that, roughly speaking, the Majorana phases must sum to a multiple of $\pi$. Of course, these phases are important with regards to $CP$-violating processes but this should suggest that their individual values should be multiples of $\pi$, not their sum. For the time being we leave this puzzle as a comment to be understood in greater detail. 

\begin{figure}
    \centering
    \includegraphics[scale = 0.43]{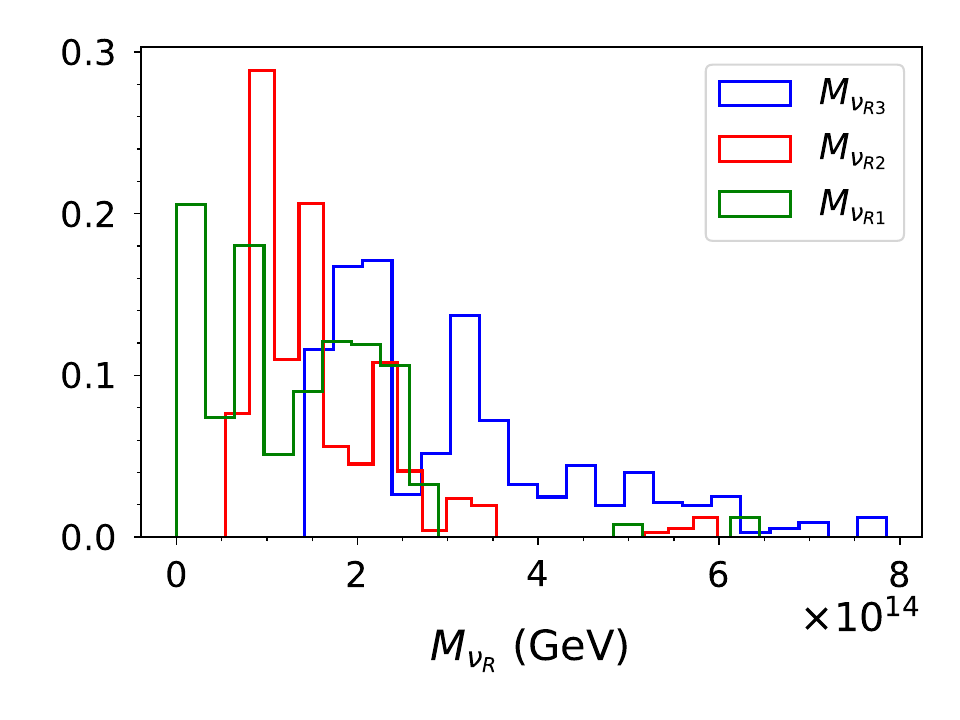} 
    \caption{Mass distribution of right handed neutrinos. Overall, the masses are very large and approximately of the same order of magnitude ($10^{14}$ GeV).}
    \label{fig:RHv}
\end{figure}

\subsection{SUSY spectrum}
\label{Sec:SUSYspectrum}

\begin{figure}
    \centering
    \includegraphics[scale = 0.43]{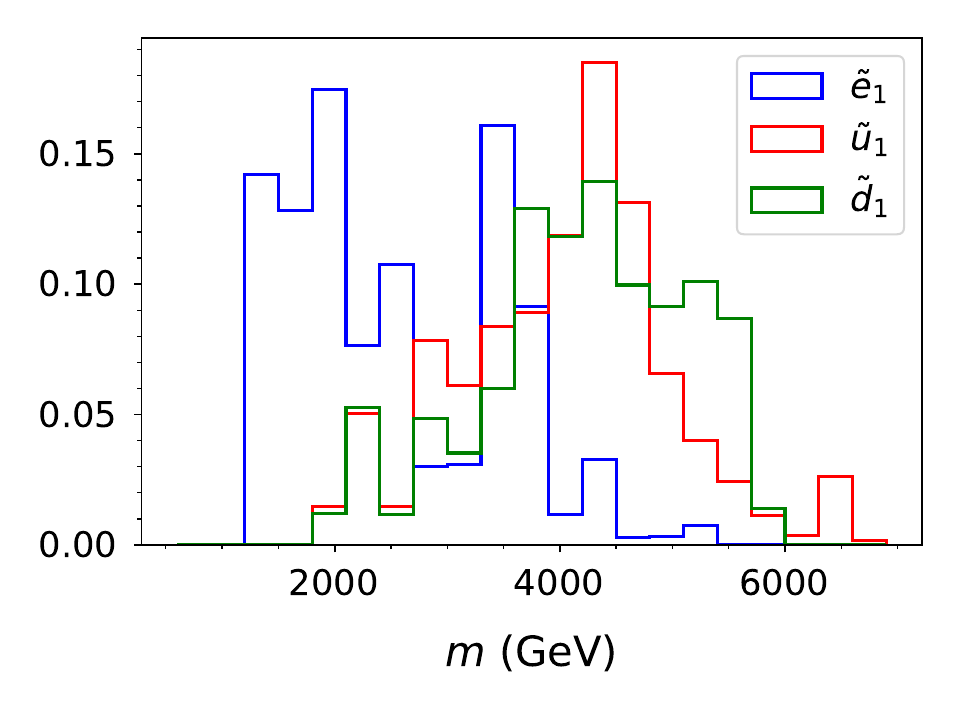} \ \ \includegraphics[scale = 0.43]{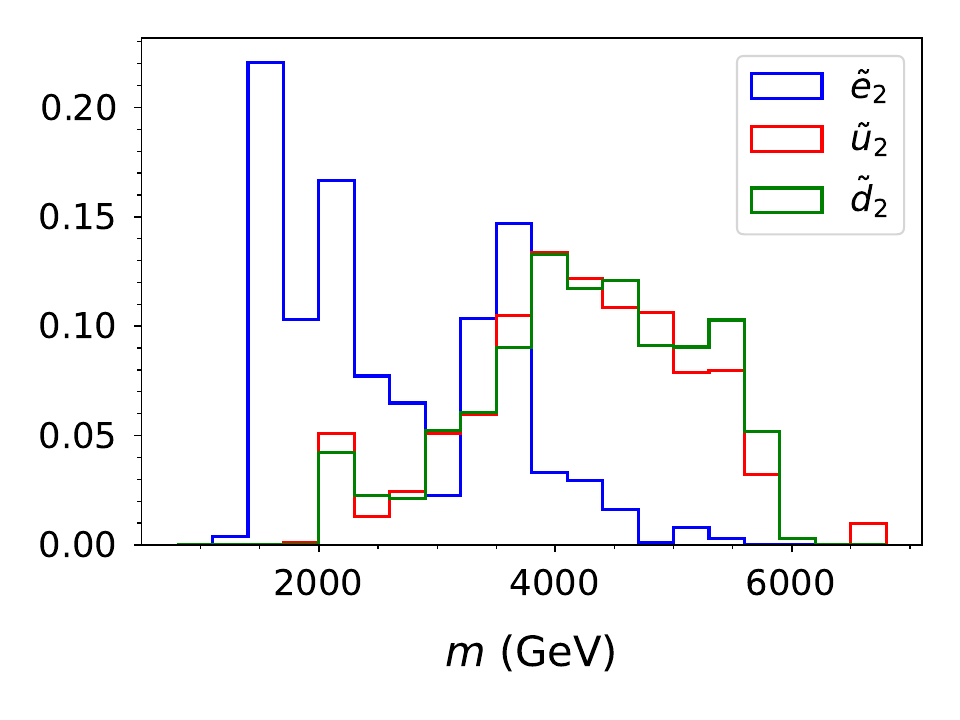}
    \caption{Distribution of masses for the lightest up-type, down-type  sfermions and the lightest slepton. The slepton is the lightest of these particles.  }
    \label{fig:1st gen masses}
\end{figure}

Figure \ref{fig:1st gen masses} shows the distribution of the lightest charged sfermion masses. We recall from table~\ref{tab:field_content} that in the sfermion sector the $SU(5)$ 5-plets, the first two generation 10-plets and the third generation 10-plet transform as a triplet, doublet and singlet under $S_4$, respectively. The nature of the lightest sfermions depens on the mass hierarchy of these states at the GUT scale. In case that $S_4$ doublet is the lightest one, then the lightest sfermion consists of a strong admixture of the first two generation sfermions. More precisely, the lightest slepton, $d$-type squark and $u$-type squark will be an admixture of $\tilde e_R$ and $\tilde \mu_R$, $\tilde d_L$ and $\tilde s_L$, and $\tilde u_R$ and $\tilde c_R$, respectively. In all other cases these states will be sfermions of the third generation. On average the corresponding mass splitting is larger in these cases compared to the first one. This can be traced back to the large top Yukawa coupling as well as to a possible difference between the masses of the 10-plet and the 5-plet at the GUT scale.

\begin{figure}
    \centering
    \includegraphics[scale = 0.43]{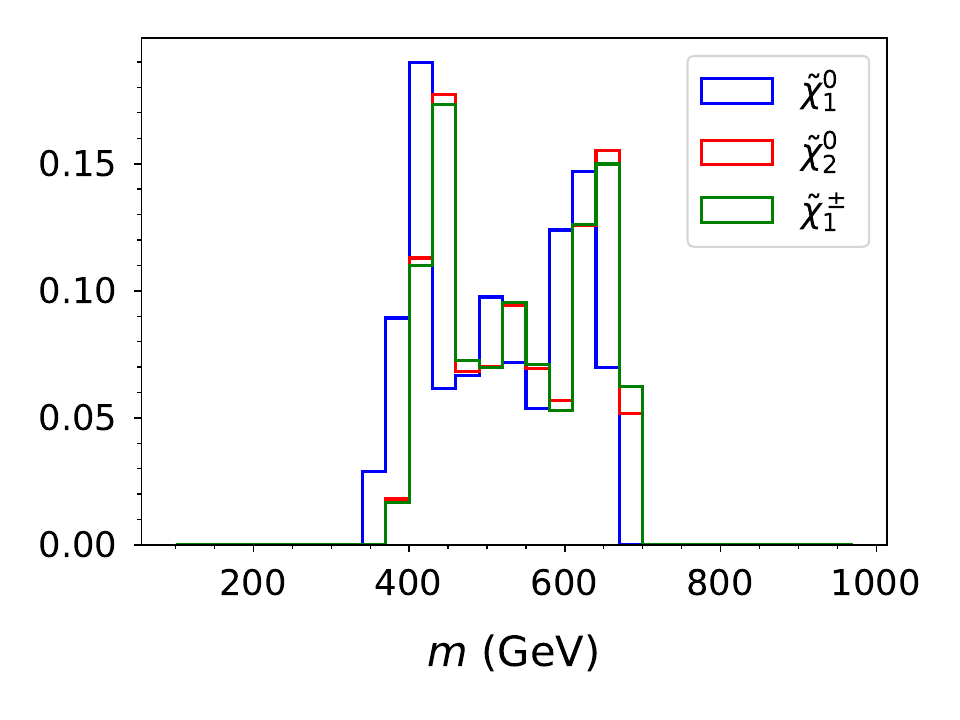}  \includegraphics[scale = 0.43]{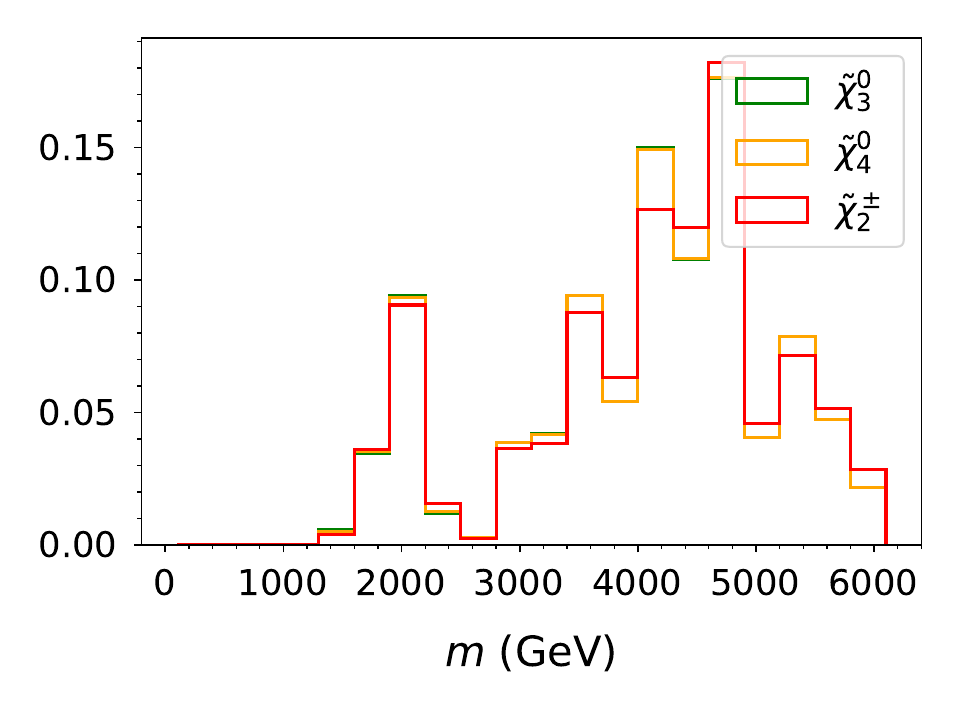} \\ 
    \includegraphics[scale = 0.43]{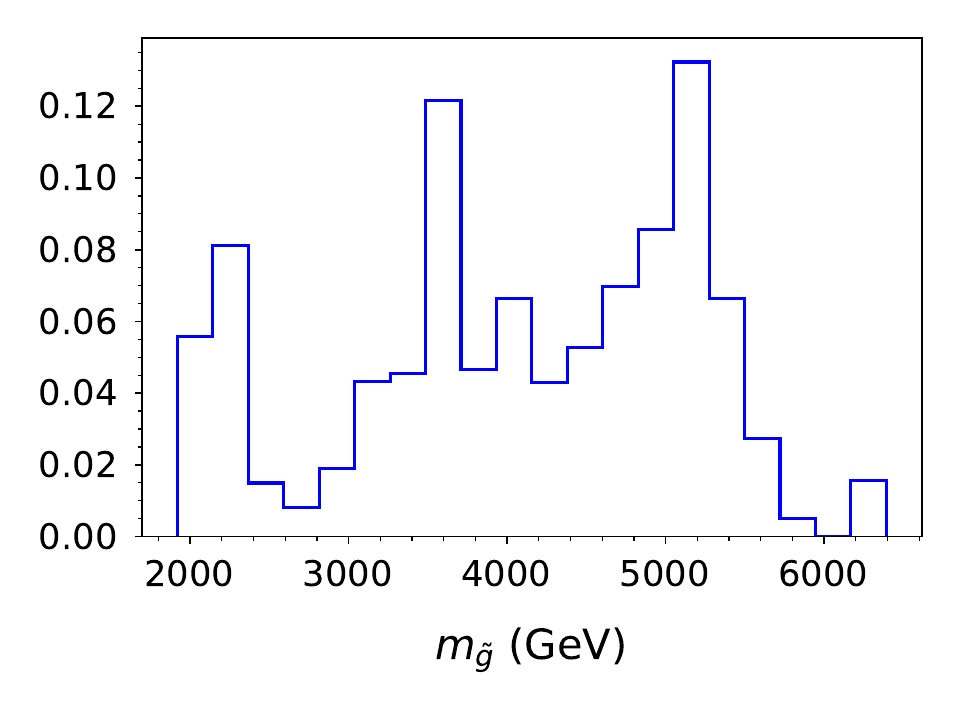}
    \includegraphics[scale = 0.43]{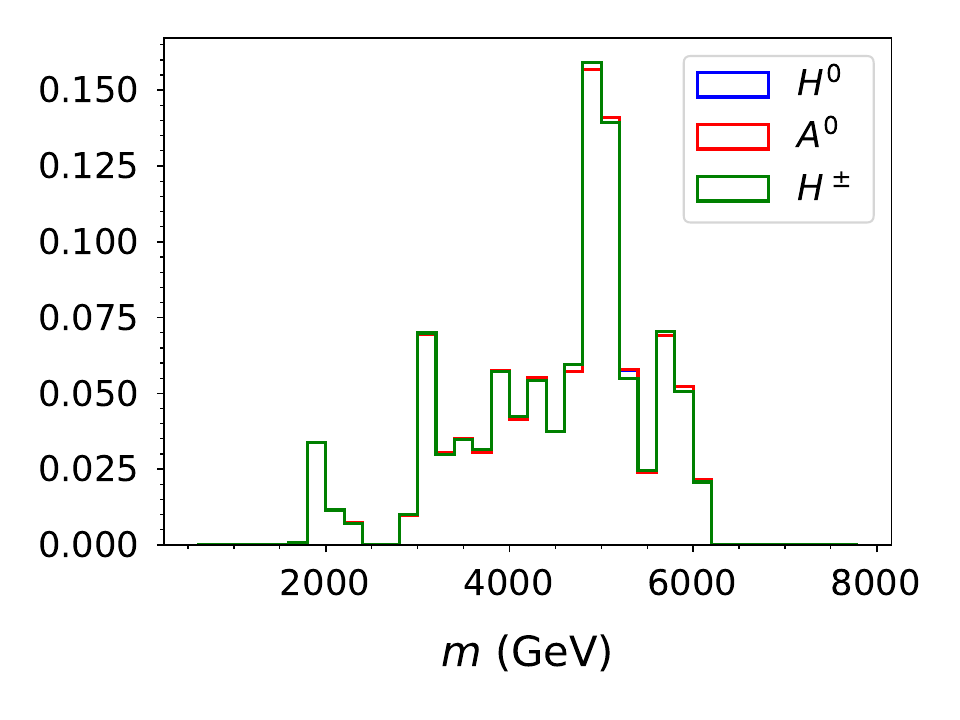}  
    \caption{Distribution of the masses of the gauginos and higgses. The mass spectrum of the lightest chargino and two lightest neutralinos is compressed to provide co-annihilation mechanism for dark matter. }
    \label{fig:gauginos}
\end{figure}

The neutralino, chargino, and gluino masses are displayed in Figure \ref{fig:gauginos}. Although the gluino mass is not particularly constrained, the two lightest neutralinos and the lightest chargino have a very constrained spectrum. As, a priori, the relic density is too high, the model requires a specific mechanism to reduce the dark matter relic density to phenomenological values. As much of the rest of the spectrum is large, the neutralinos and charginos supply an alternative mechanism via co-annihilation. In order to allow for such contributions, the lightest gauginos must be comparable in mass as will be seen in the dark matter dedicated section below. 

The last panel of Figure \ref{fig:gauginos} demonstrates the "decoupling limit" for two Higgs doublet models. In this limit, the three additional Higgs states have very large masses and are approximately degenerate. 

Finally, Figure \ref{fig:RHv} depicts the mass distribution of the heavy neutrinos. Overall, the masses are very large and approximately of the same order of magnitude ($10^{14}$ GeV).

\subsection{Dark matter}

\begin{figure}
    \centering
    \includegraphics[scale = 0.43]{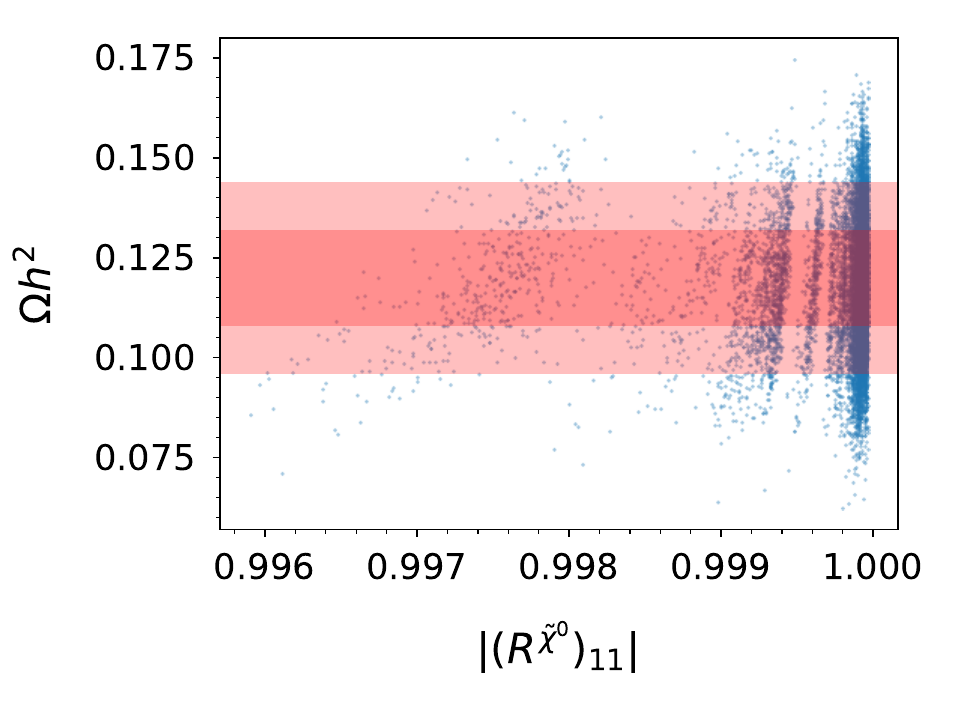} 
    \includegraphics[scale = 0.43]{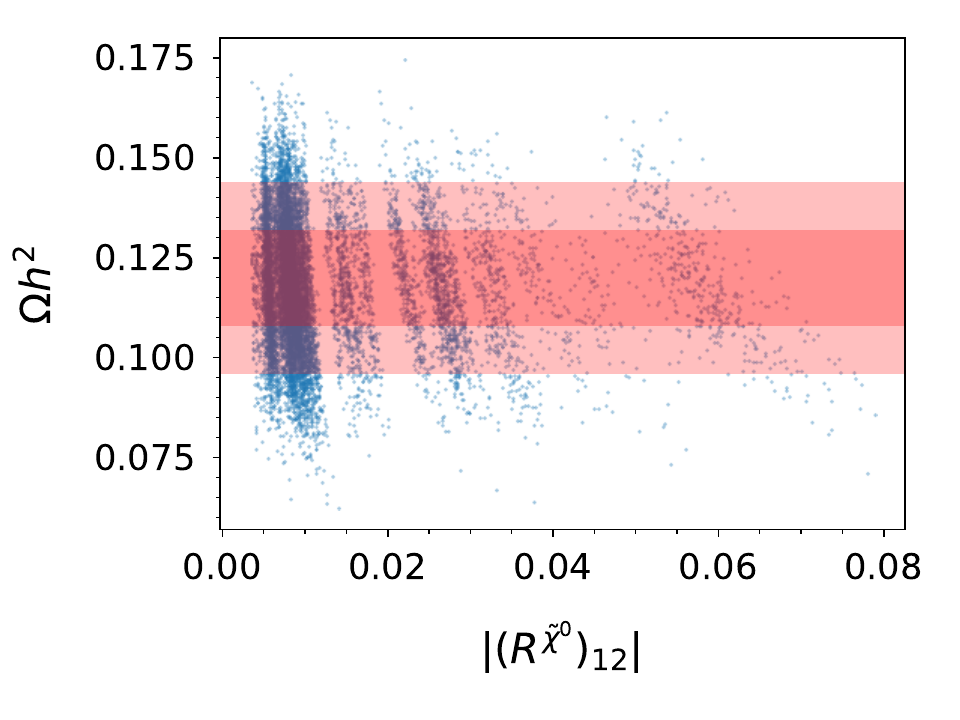} 
    \caption{The bino ($|(R^{\tilde{\chi}^0})_{11}|$, left) and wino ($|(R^{\tilde{\chi}^0})_{12}|$, right) contents of the lightest neutrino shown against the corresponding dark matter relic density. The higgsino contents of the lightest neutralino are negligible and not shown here. The neutralino mixing matrix $R^{\tilde{\chi}^0}$ is defined according to the SLHA standard \cite{Skands:2003cj}. The red (light red) region is indicating the $1\sigma$ ($2\sigma$) limits}
    \label{fig:omegaDMvsNMIX11}
\end{figure}

We now come to the discussion of dark matter aspects of the model under consideration. As we have seen in Figure \ref{fig:example constraints}, the dark matter relic density given by the latest Planck results is well accommodated for in the parameter regions surviving the numerous imposed constraints. The corresponding parameter configurations feature essentially bino-like dark matter, which can be understood from Figure\ \ref{fig:omegaDMvsNMIX11}, where we depict the relevant bino and wino content of the lightest neutralino.

Looking at the gaugino masses (Figure\ \ref{fig:gauginos}), it can be seen that the second-lightest neutralino as well as the lighter chargino lie very close to the lightest neutralino. In other words, the bino and wino mass parameters are, at the SUSY scale, almost equal, the bino lying just below the wino mass. This feature is driven by co-annihilations needed to achieve the required relic density. Note that this corresponds to a situation where the GUT-scale values of the bino and wino mass differ roughly by a factor of two.

It is interesting to note that, although the bino-like lightest neutralino $\tilde{\chi}^0_1$ is the dark matter candidate, the (co-)annihilation cross-section is dominated by the (co-)annihilaton of the wino-like states $\tilde{\chi}^0_2$ and $\tilde{\chi}^{\pm}$. This is explained, on the one hand, by the very small mass difference between the wino-like states, and, on the other hand, by the enhanced annihilation cross-section for the latter as compared to the bino-like state. Typical final states of these (co-)annihilation channels are quark-antiquark pairs and gauge boson pairs (including $Z^0$, $W^{\pm}$, and $\gamma$). Neutrino final states are subdominant.

\begin{figure}
    \centering
    \includegraphics[scale = 0.45]{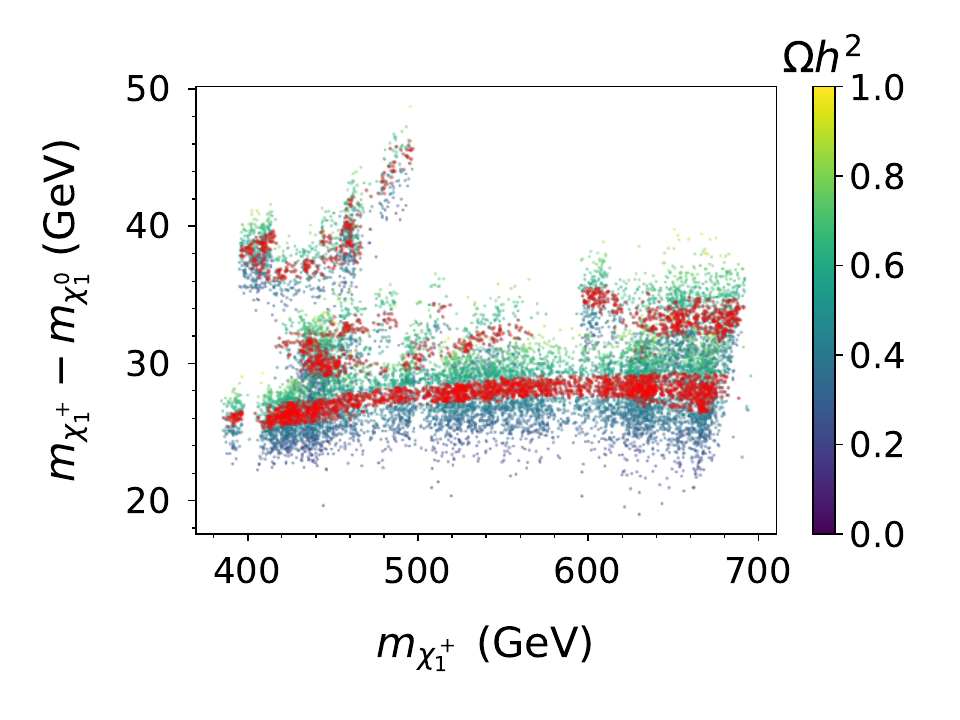}  \includegraphics[scale = 0.45]{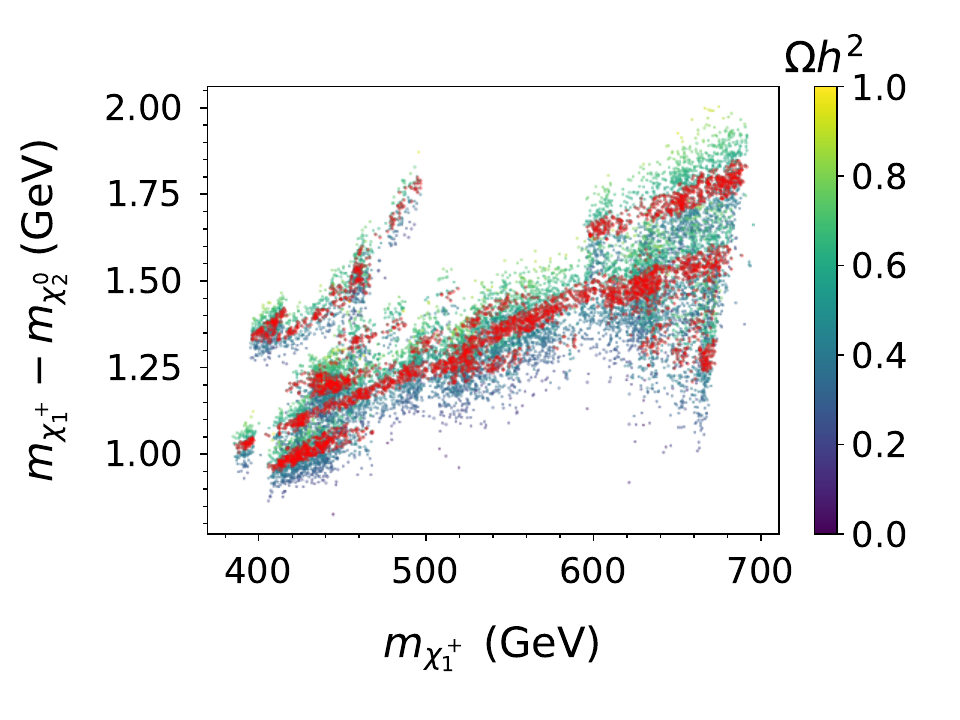} 
    \caption{Mass differences between the lightest neutralino (left) as well as the second-lightest neutralino (right) and the lightest neutralino displayed against the lightest chargino mass. Correlations between the masses are displayed with the colour indicating the relic density. In addition, red dots correspond to a relic density of $\Omega h^2 = 0.120 \pm 0.006$. The relic density is controlled by co-annihilation channels whose strength are dictated by the mass gap between the relevant particles.}
    \label{fig:gauginocorrelations}
\end{figure}

The presence of co-annihilation can also be understood through Figure\ \ref{fig:gauginocorrelations} showing the correlation of the three lightest gaugino masses and the dark matter relic density. Let us finally note that scenarios with wino-like dark matter would give rise to insufficient relic density to align with the experimental evidence, as the wino (co-)annihilation cross-section is numerically more important as the one for the bino.

\begin{figure}
    \centering
    \includegraphics[scale = 0.43]{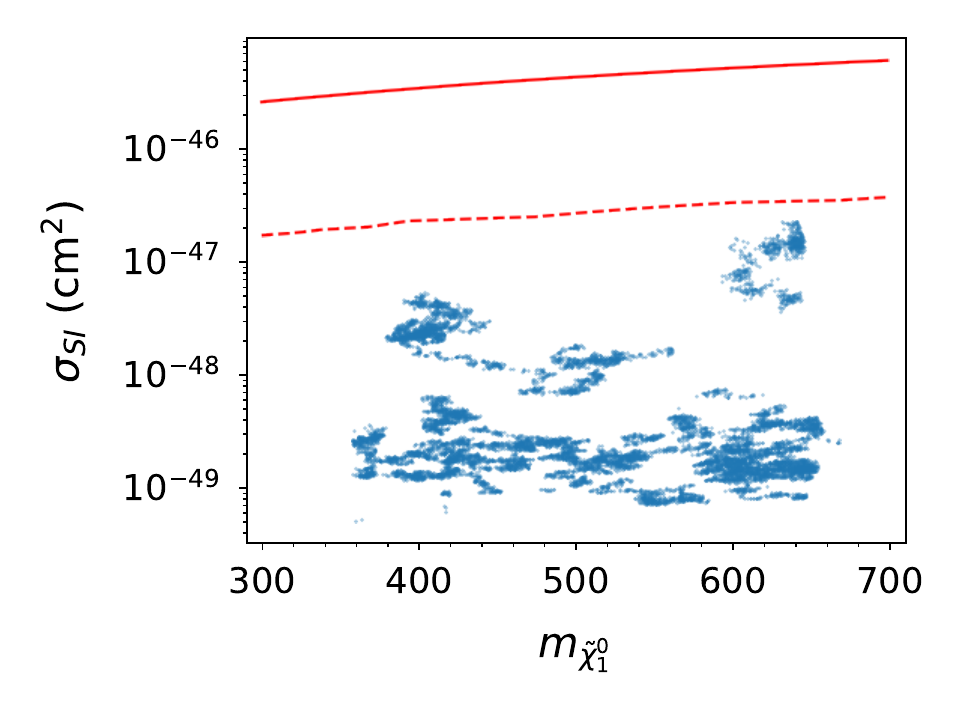}\ \ 
    \includegraphics[scale = 0.43]{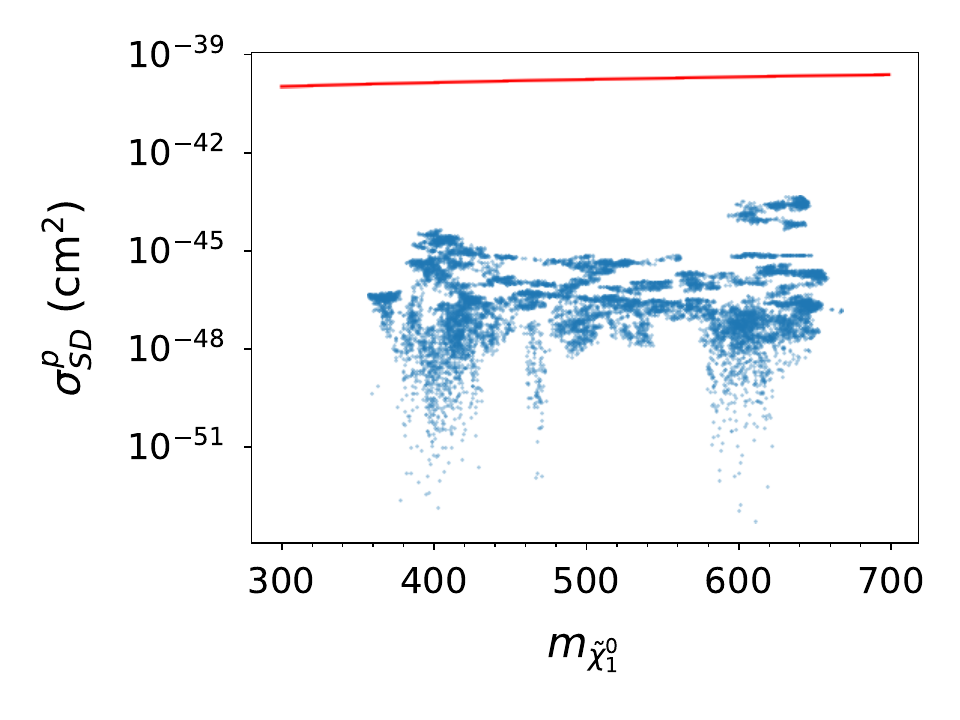} \ \ 
    \caption{The direct detection limits for spin dependant and spin independent cross-sections are shown with the experimental limit plotted. The solid line indicates the {\tt XENON1T} limits \cite{Aprile:2019dbj}, while the dashed line in the first panel indicates the expected limit for the {\tt XENONnT} \cite{XENON:2020kmp} experiment. We have precluded a representation of the neutron direct detection calculation as all data points are far away from the exclusion limit, much like the proton calculation of the same.}
    \label{fig:DDSI}
\end{figure}

Coming to the direct dark matter detection, we can see from Figure\ \ref{fig:DDSI} that this constraint is also well satisfied in the model under consideration, both for the spin-dependent and the spin-independent case. It is important to note that all points shown in Figure\ \ref{fig:DDSI} lie also below the projected limits of the {\tt XENONnT} experiment \cite{XENON:2020kmp}. The fact that all points are found below this limit can be traced to the fact that we have applied a cut on the global likelihood value as explained in Section\ \ref{sec:method}. This procedure discards the points which are too close to the current {\tt XENON1T} limit, since they typically feature a somewhat lower likelihood value. This means that parameter configurations with reasonably high global likelihood values may not be challenged by direct detection experiments in a near future.

In summary, the relic density constraint implies a relatively small mass difference between the lightest neutralino and the next-to-lightest states, leading to final states with soft pions and leptons which are difficult to detect. The current bounds depend on the nature of the NLSP go up to masses of about 240 GeV \cite{ATLAS:2017vat, CMS:2018kag, CMS:2019san, ATLAS:2019lng}. This cuts slightly into the allowed parameter space.

\subsection{Collider related aspects}

\begin{figure}
    \centering
    \includegraphics[scale = 0.43]{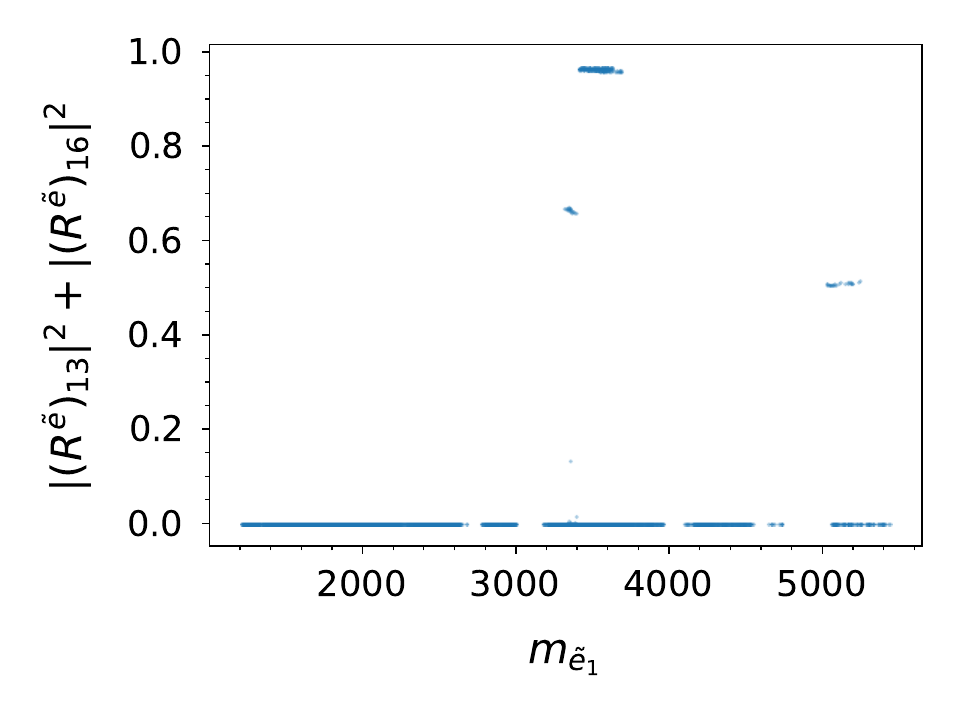} \ \ \includegraphics[scale = 0.43]{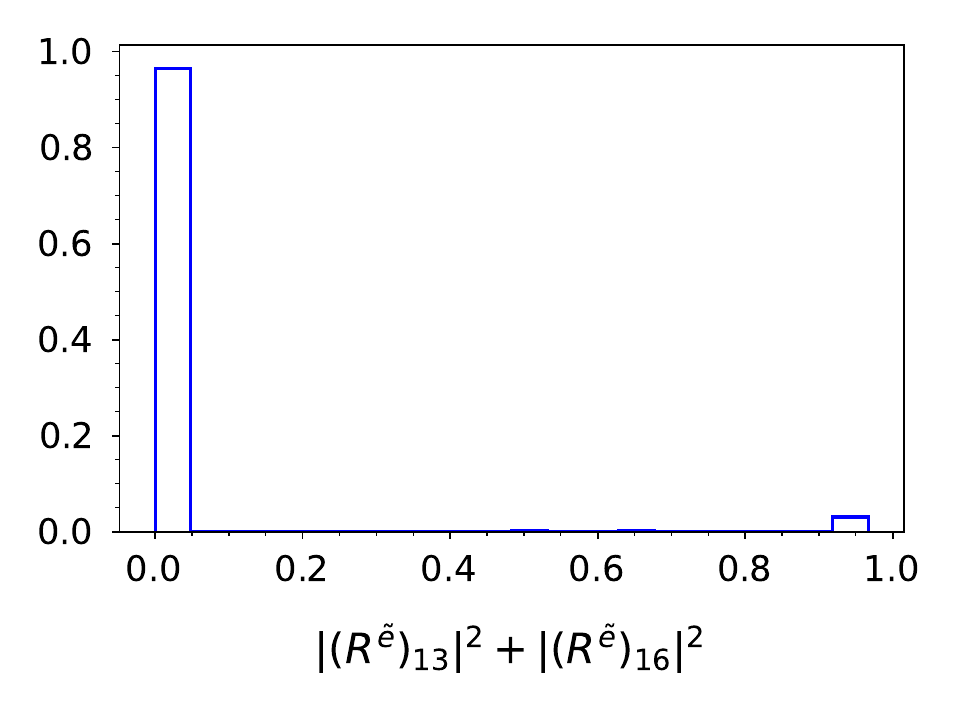} 
    \\
    \includegraphics[scale = 0.43]{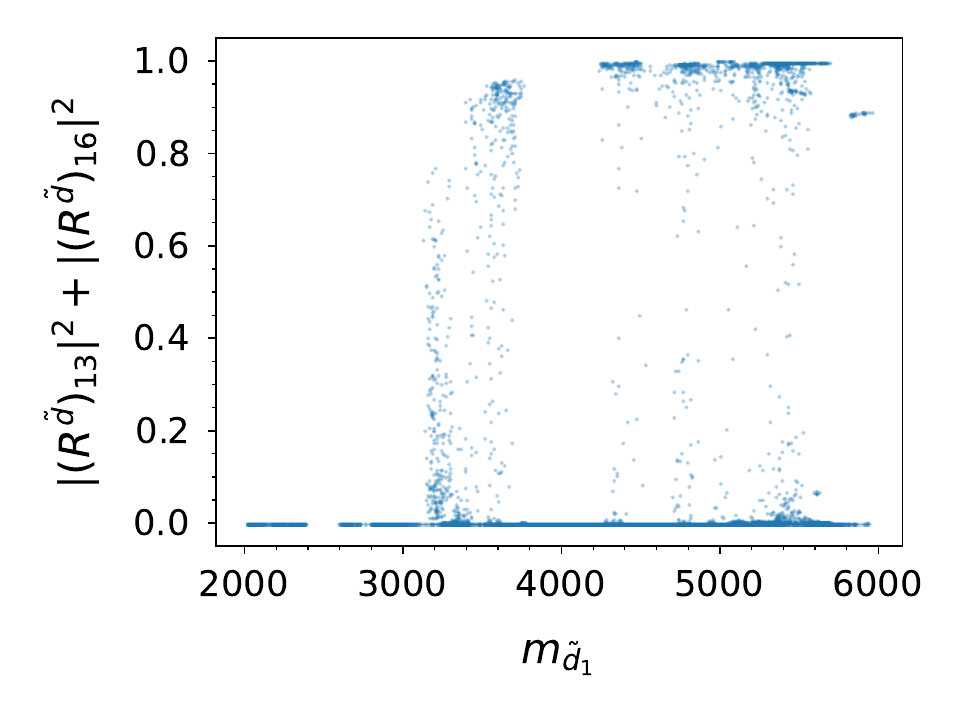} \ \ \includegraphics[scale = 0.43]{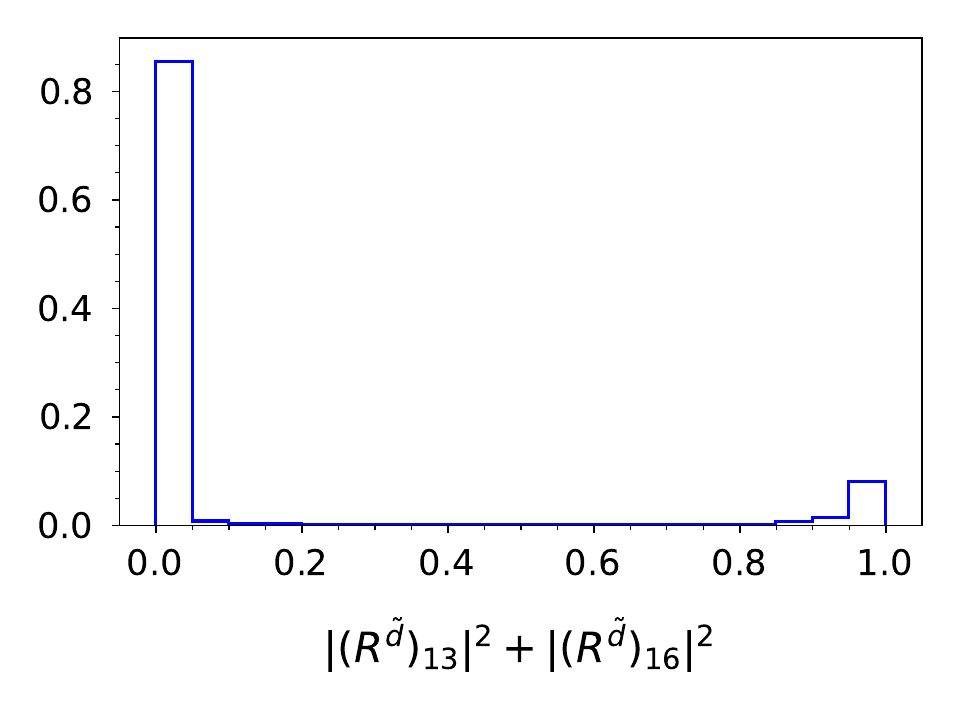} 
    \\
    \includegraphics[scale = 0.43]{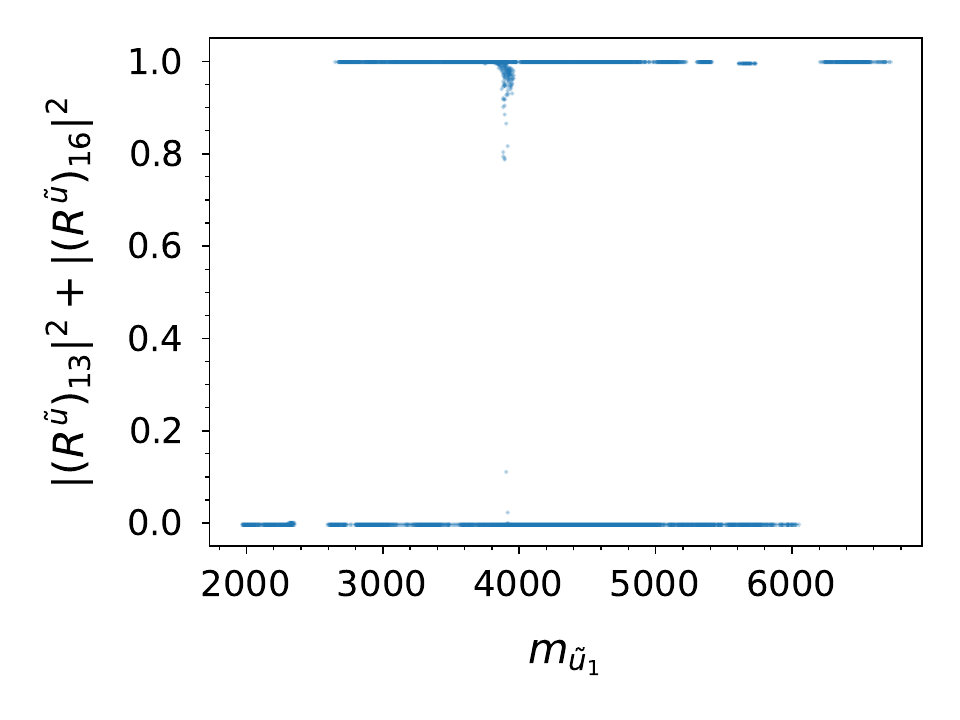} \ \ \includegraphics[scale = 0.43]{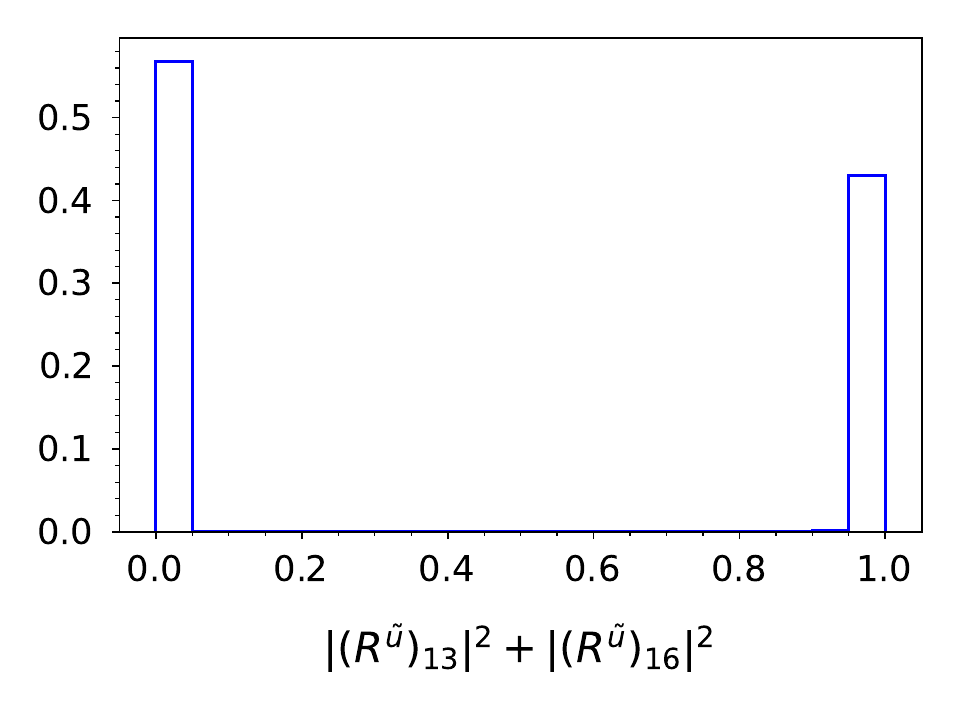} 
    
    \caption{The mass of the lightest sparticle for up and down type squarks and sleptons is plotted against the third generation content for the given particle on the left panels. When the third generation content is close to 0, the sfermion features a second and first generation maximally mixed state. For low mass down squark and slepton states, an admixture of first and second generation is favoured. The right panels illustrate the actual proportion of points that belong to these two extreme flavour cases. The sfermion mixing matrices $R^{\tilde{f}}$ ($\tilde{f} = \tilde{u}, \tilde{d}, \tilde{e}$) are defined according to the SLHA 2 standard \cite{Allanach:2008qq}.}
    \label{fig:sfermionflavourcompo}
\end{figure}

As already mentioned in Section\ \ref{Sec:SUSYspectrum} the flavour structure of the lightest sfermions falls into two extreme case: For each sfermion type, we observe either a first and second generation fully mixed state or a strict third generation state. This feature is illustrated in Figure \ref{fig:sfermionflavourcompo} where we show the distribution of the sum of the square of the mixing matrix entries. 

While at first glance it seems that this particular prediction of the model would be quite interesting from a collider perspective, enabling potential flavour mixed search channels, the model also predicts rather high masses for sfermion states. The lightest squark masses are peaked around 4 TeV and stand well beyond any potential collider sensitivity reach (see for e.g.\ Ref.\ \cite{ATLAS:2020syg} where the limits are around 1.8 TeV squarks using simplifying assumptions that do not hold in our case). The lightest slepton on the other hand can be as light as 1 TeV. However, the production cross section drops significantly with respect to the QCD dominated squark ones. Furthermore, our model naturally predicts right-handed slepton states to be the lightest ones which further decreases the production cross section
\cite{Beenakker:1999xh}. As an example, recent searches for flavour conserving channels with nearly helicity degenerated slepton states are excluding masses of the order 600 -- 700 GeV \cite{ATLAS:2019lff} if $m_{\tilde \chi^0_1} < 150$ GeV (and limits are even weaker if this is not the case). In addition, current exclusion on $\tau$ flavoured sleptons masses are of order 390 GeV \cite{ATLAS:2019gti}. 

\begin{figure}
    \centering
    \includegraphics[scale = 0.49]{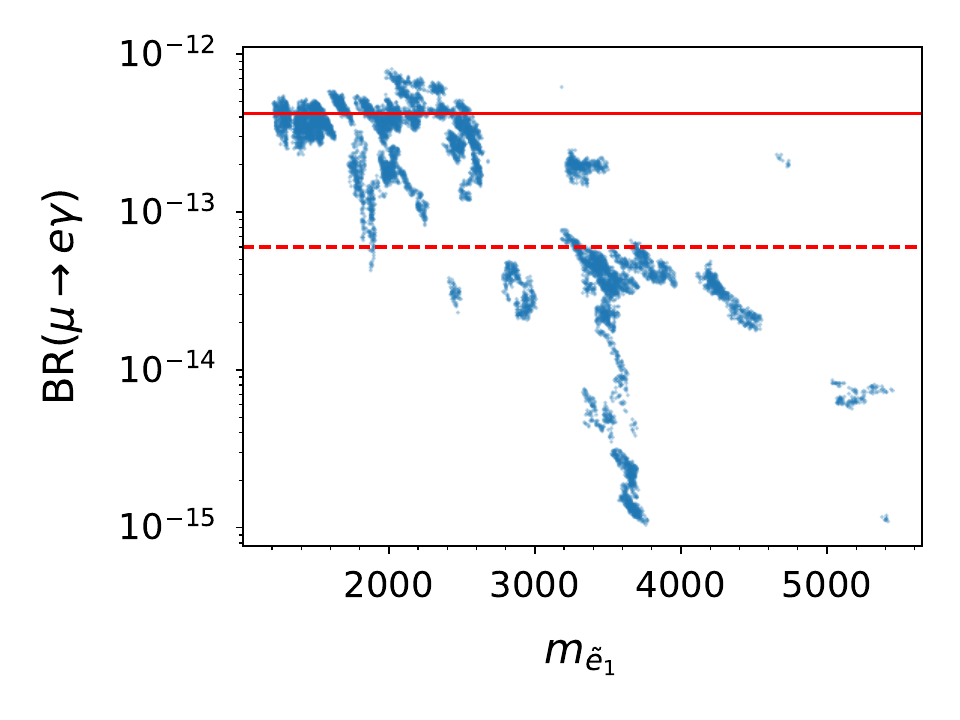}
    \caption{BR$(\mu \rightarrow e \gamma)$ as a function of $m_{\tilde{e}_1}$. Current experimental limit is represented by the red solid line while the dashed one corresponds to the future prospects of MEG II ($6 \cdot 10^{-14}$) \cite{MEGII:2018kmf}.}
    \label{fig:mu_egamma_mse1}
\end{figure}

It turns out that our model predictions regarding the sfermionic sector are not very promising for potential collider searches. However this flavour mixing particularity leads to some interesting features from indirect searches perspectives. It is well known that slepton mixing can generate significant contributions to flavour violating decay constraints; in particular, if the mass of the slepton is rather light. In our case, BR$(\mu \rightarrow e\gamma)$ illustrates very well this feature as being one of the most stringent test for lepton flavour violation. Figure \ref{fig:mu_egamma_mse1} shows the distribution of this branching ratio as a function of the mass of the lightest slepton. While the points are within the current experimental limits, it appears that future prospects in the current MEG II experiment \cite{MEGII:2018kmf} could rule out a vast majority of our light slepton points, implying the interesting conclusion that this particular constraint and other flavour violating constraints would have more discriminatory power than classic direct slepton searches and propose typical smoking guns for our framework. We also note that, while we can have light staus in the model, $\tau \rightarrow e\gamma$ limits are far from constraining our model with branching ratios below $10^{-13}$ while current limits are around $10^{-8}$.

Despite the sfermionic states being unreachable in near future collider searches, this is not necessarily the case for the other SUSY particles in our model. In particular, the model can predict rather light electroweakinos with sufficient mass gap for collider considerations. As a comparison, recent searches from ATLAS \cite{ATLAS:2021moa} put a lower bound on $m_{\tilde{\chi}_{1}^0}$ of 270 GeV when the mass gap with $m_{\tilde{\chi}_{1}^{\pm}}$ is of order 50 GeV. While this very light masses are not present in our current framework, we can hope for more stringent limits from future LHC runs.

Similar conclusion can be derived regarding gluino searches. While the predicted masses lie in the 2 -- 6 TeV range, ATLAS and CMS limits using simplified models \cite{ATLAS:2020syg, CMS:2019zmd} reaches 1.8 TeV exclusion. Again, we can argue that future LHC runs might restrict our light spectrum parameter distributions.

To illustrate the collider phenomenology we present three benchmark points where relevant information is given in Table \ref{Tab:BPs}. The detailed information is added as supplementary material to the arXiv submission of this paper. In particular, we list the masses of the electroweakinos for an illustrative point with low masses for the light gauginos and a relatively small mass gap. We also give the dominant decay channels and decay widths. All other particles are too heavy to be detected in the upcoming LHC run. We give a benchmark for a point which features, besides the light chargino and the neutralinos of BP1, a gluino with mass 2 TeV which should be in reach of the upcoming LHC run. Again, we give the masses, decay widths, and decay channels for the particles under examination. Regarding the third benchmark point, while having similar features as BP2 in view of collider physics, it further highlights the fully mixed nature of the lightest sleptons and the potential relevance for $\mu \to e \gamma$. Its branching of $4\cdot 10^{-13}$ is close to the current experimental bound. However, even if this rare decay is discovered in an upcoming experiment, this example shows that it will be rather challenging to detect lepton flavour at a high energy collider as the sleptons are quite heavy.

\begin{table}
\centering
\begin{tabular}{|c|c|c|c|c|c|}
\hline
                     & Particle           & Mass   & Decay Width          & Decay Channels                                                                                                                              & Branching Ratio                                            \\ \hline \hline
\multirow{3}{*}{BP1} & $\tilde{\chi}_1$   & $402$  & $0$                  &                                                                                                                                             &                                                            \\ \cline{2-6} 
                     & $\tilde{\chi}_2$   & $442$  & $1.76\times10^{-11}$ & \begin{tabular}[c]{@{}c@{}}$\tilde\chi_1 \:\bar{q}\:q $\\ $\tilde\chi_1 \:\bar{b}\:b $\\ $\tilde\chi_1 \: \nu_i \nu_i  $\end{tabular}       & \begin{tabular}[c]{@{}c@{}}70\%\\ 13\%\\ 14\%\end{tabular} \\ \cline{2-6} 
                     & $\tilde{\chi}_1^+$ & $444$  & $1.5\times10^{-8}$   & \begin{tabular}[c]{@{}c@{}}$\tilde\chi_1 \:\bar{q}\:q $\\ $\tilde\chi_1 \:\bar{l}\:\nu_i $\end{tabular}                                     & \begin{tabular}[c]{@{}c@{}}67\%\\ 33\%\end{tabular}        \\ \hline \hline
BP2                  & $\tilde {G}$       & $2000$ & $1.87\times10^{-1}$  & \begin{tabular}[c]{@{}c@{}}$\tilde{\chi_1} \:\bar{q}\:q $\\ $\tilde{\chi_1^{\pm}} \: q^{*} \:q $ \\ $\tilde{\chi}_2 \: q^{*} \:q $\end{tabular} & \begin{tabular}[c]{@{}c@{}}11\%\\ 46\%\\ 35\%\end{tabular} \\ \hline \hline
\multirow{2}{*}{BP3} & $\tilde{e}_1$      & $1470$ & $6.6$                & \begin{tabular}[c]{@{}c@{}}$\tilde\chi_1 \:e $\\ $\tilde\chi_1 \:\mu $\end{tabular}                                                         & \begin{tabular}[c]{@{}c@{}}44\%\\ 56\%\end{tabular}        \\ \cline{2-6} 
                     & $\tilde{e}_2$      & $1500$ & $6.8$                & \begin{tabular}[c]{@{}c@{}}$\tilde\chi_1 \:e $\\ $\tilde\chi_1 \:\mu $\end{tabular}                                                         & \begin{tabular}[c]{@{}c@{}}56\%\\ 44\%\end{tabular}        \\ \cline{2-6} 
                     & $\tilde{\chi}_1$	& $413$ & $0$					&                                                                                                                                             &                                                            \\  \cline{2-6} 
                     & $\tilde{\chi}_2$	& $439$ & $3.72\times10^{-11}$	& \begin{tabular}[c]{@{}c@{}}$\tilde\chi_1 \:q^* \:q $\\ $\tilde\chi_1 \:\nu \: \nu $\end{tabular} 		& 	\begin{tabular}[c]{@{}c@{}}81\%\\ 15\%\end{tabular} \\  \cline{2-6} 
                     & $\tilde{\chi}_1^+$ &$440$ & $5.18\times10^{-8}$	& \begin{tabular}[c]{@{}c@{}}$\tilde\chi_1 \:q^* \:q $\\ $\tilde\chi_1 \:\nu \:e^* $\end{tabular} 		& 	\begin{tabular}[c]{@{}c@{}}67\%\\ 33\%\end{tabular} \\  \cline{2-6} 
                     & $\tilde {G}$ 	& $2110$ & $0.228$				& \begin{tabular}[c]{@{}c@{}}$\tilde\chi_1 \:q^* \:q $\\ $\tilde\chi_1^\pm \:q^* \:q $\\ $\tilde\chi_2 \:q^* \:q $\end{tabular} 		& 	\begin{tabular}[c]{@{}c@{}}13\%\\ 50\%\\37\%\end{tabular}  \\ \hline
\end{tabular}
\caption{Selected benchmark points (BP) for phenomenology. The first BP exhibits rather light electroweakinos and might represent a challenge for future colliders. The second BP has a rather light gluino, which is already close to the current experimental limits from the LHC collaborations. Finally, the last BP is an example of maximally mixed lightest sleptons. All masses and widths are given in GeV.}
\label{Tab:BPs}
\end{table}
\section{Conclusion}
\label{sec:conc}

In this paper we have presented a detailed phenomenological analysis of a concrete Supersymmetric (SUSY) Grand Unified Theory (GUT) of flavour, based on $SU(5)\times S_4$. The model predicts charged fermion and neutrino mass and mixing, and where the mass matrices of both the Standard Model and the Supersymmetric particles are controlled by a common symmetry at the GUT scale, with only two input phases. The considered framework predicts small but non-vanishing non-minimal flavour violating effects, motivating a sophisticated data-driven parameter analysis to uncover the signatures and viability of the model. 

The computer-intensive Markov-Chain-Monte-Carlo (MCMC) based analysis performed here, the first of its kind to include a large range of flavour as well as dark matter and SUSY observables, predicts distributions for a range of physical quantities which may be used to test the model. The predictions include maximally mixed sfermions, $\mu\rightarrow e \gamma$ close to its experimental limit and successful bino-like dark matter with nearby winos (making direct detection unlikely), implying good prospects for discovering winos and gluinos at forthcoming collider runs. The results also demonstrate that the Georgi-Jarlskog mechanism does not provide a good description of the splitting of down type quark masses and charged leptons. However neutrinoless double beta decay, which depends on a curious pattern of Majorana phases resulting from the two input phases, is predicted at observable rates.

The analysis here may be repeated for any given SUSY GUT of flavour, leading to corresponding predictions for fermion masses and mixing as well as SUSY masses and flavour violating physical observables at colliders and high precision experiments. The results here exemplify the synergy between the theory of quark and lepton (including neutrino) mass and mixing, dark matter and the SUSY particle spectrum and flavour violation, that is possible within such frameworks. It is only by systematically confronting the detailed predictions of concrete examples of SUSY GUTs of flavour with experiment that the underlying unified theory of quark and lepton flavour beyond the Standard Model may eventually be discovered.

\acknowledgments
S.\,J.\,R.\, and A.\,K.\,F.\, are supported by Mayflower studentships from the University of Southampton. This work is supported by {\it Investissements d’avenir}, Labex ENIGMASS, contrat ANR-11-LABX-0012. S.F.K. acknowledges the STFC Consolidated Grant ST/L000296/1 and the European Union's Horizon 2020 Research and Innovation programme under Marie Sk\l{}odowska-Curie grant agreement HIDDeN European ITN project (H2020-MSCA-ITN-2019//860881-HIDDeN. This research
was supported by the Deutsche Forschungsgemeinschaft (DFG, German Research Foundation) under grant 396021762 - TRR 257.
\appendix
\section{SPheno}
\label{app:spheno}

Our results have been obtained using the numerical code {\tt SPheno} \cite{Porod:2003um,Porod:2011nf}, where we have implemented the model using \texttt{SARAH v4.14.0} \cite{Staub:2008uz,Staub:2013tta,Staub:2012pb,Staub:2010jh,Staub:2009bi}
and then adapted the code for the model at hand. Here we describe the corresponding modifications.

In the standard version of \texttt{SPheno} the SM fermion masses, the CKM matrix, the mass of the $Z$-boson $m_Z$, the Fermi constant, the electromagnetic coupling $\alpha$, and the strong coupling $\alpha_s(m_Z)$ serve as input.
The latter can either be given in the Thomson limit or in the $\overline{\rm MS}$-scheme at $Q=m_Z$. From these the three gauge couplings, the SM vacuum expectation value (VEV) $v$ of the Higgs boson and the Yukawa couplings are calculated at the scale $Q=m_Z$. The couplings are then evolved up to the scale $M_{\rm SUSY}$ where the SM and the MSSM are matched including one-loop SUSY threshold corrections.

In the model at hand the Yukawa couplings are given at the GUT-scale and the fermion masses, the CKM-matrix and the PMNS-matrix are an output which required some changes to the code. The input which is done via the standard SUSY Les Houches format \cite{Skands:2003cj, Allanach:2008qq} with the slight modification that the Yukawa couplings can be input at the GUT scale.
The input is given at different scales as follows:
 \begin{itemize}
   \item at $M_{\rm GUT}$: $Y_{\ell}$, $Y_\nu$, $Y_d$, $Y_u$, arg($\mu$)
     (in practice the sign of $\mu$), as well as
     the soft SUSY breking terms in a non-universal form: scalar mass squares $m^2_{\tilde f}$ ($\tilde f=\dots$), trilinear couplings $A_{\tilde f}$ ($\tilde f=\dots$) and non-universal gaugino mass parameters $M_1$, $M_2$, $M_3$. All phases can be non-zero in principle. 
   \item at $M_{\rm SUSY}$: $\tan\beta$
   \item at $Q=m_Z$: $G_F$, $m_Z$, $\alpha_s$,  $\alpha_{em}(m_Z)$
 \end{itemize}
The calculation is done in an iterative way: 
 \begin{enumerate}
 \item The gauge couplings are evolved from
  the electroweak scale using the SUSY RGEs at the one-loop level to
  $M_{\rm GUT} = 2 \cdot 10^{16}$~GeV. We do not require
  unification at this scale. These couplings at this stage serve only as a starting point for our iteration.
 \item All parameters are evolved from $M_{\rm GUT}$ to $M_{\rm SUSY} = \sqrt{ (M_Q)_{33} (M_U)_{33} }$
 using RGEs at the two-loop level.
 The right-handed neutrinos are decoupled at their
 respective mass scale during this evaluation and
 the contributions to the Weinberg operator are calculated at these scales as well. The running of this operator is taken into account as well. 
 \item The SUSY spectrum is calculated at the
 scale $M_{\rm SUSY}$ at the one-loop level and the heavy Higgs masses at two-loop level taking into account the
 contributions from third generation sfermions and fermions, see Ref.~\cite{Slavich:2020zjv} for a summary. In addition the matching to the SM-parameters is performed as described in Ref.\ \cite{Staub:2017jnp}.
 \item The SM-parameters are evolved from $M_{\rm SUSY}$ to $m_Z$ using the two-loop SM RGEs. At $m_Z$, the masses of the SM fermions are calculated and the mass of the Higgs boson at the two-loop level. For these calculations we however take $G_F$, $m_Z$,
 $\alpha(m_Z)$ and $\alpha_s(m_Z)$ to calculate
 the gauge couplings and the VEV. 
 \item The gauge and Yukawa couplings as well as the quartic Higgs couplings are then evolved to
 $M_{\rm SUSY}$ using the two-loop RGEs. At this scale
 the SUSY threshold corrections to the gauge and Yukawa couplings are taken into account. The
 resulting couplings are evolved to $M_{\rm GUT}=2\cdot 10^{16}$ GeV. Then steps 2.\ to 4.\ are repeated until a relative precision of all masses at the level of $10^{-5}$ is reached.
 \item Once the precision goal for the spectrum has been achieved, the flavour observables are calculated. Also here we have modified the procedure slightly: we have a quite heavy spectrum leading potentially large logs of the
 form $\ln(M_{\rm SUSY}/m_t)$. For this reason the calculation is done in two steps: (i) Calculate the SUSY contributions to the Wilson coefficients at $Q=M_{\rm SUSY}$. (ii) Calculate the SM contributions to the Wilson coefficients at $Q=m_t$.
 (iii) Add both contributions to calculate the relevant observables.
 \end{enumerate}

\bibliographystyle{JHEP}
\bibliography{refs}

\end{document}